\documentclass[preprints,article,accept,oneauthor,pdftex]{Definitions/mdpi} 

\firstpage{1} 
\pubvolume{1}
\issuenum{1}
\articlenumber{0}
\pubyear{2021}
\copyrightyear{2020}
\datereceived{} 
\dateaccepted{} 
\datepublished{} 
\hreflink{https://doi.org/} 

\usepackage{newtxtext}

\usepackage{natbib}
\usepackage{color}
\usepackage[english]{babel}
\usepackage{amsmath}
\usepackage{amssymb}
\usepackage{bm}
\usepackage[utf8]{inputenc}
\usepackage{ulem}
\usepackage{hyperref}
\usepackage{amsfonts}
\newcommand{\dd}{\mathrm{d}}
\newcommand{\rhob}{\rho_\text{b}}
\newcommand{\Tb}{T_\text{b}}
\newcommand{\kB}{\mbox{$k_\text{B}$}}

\newcommand{\gcc}{\mbox{g~cm$^{-3}$}}
\newcommand{\xg}{x_\text{g}}
\newcommand{\rg}{r_\text{g}}

\newcommand{\Ts}{T_\text{s}}

\newcommand{\SB}{\sigma_{\rm SB}}
\newcommand{\Omegas}{\Omega_\text{s}}
\newcommand{\Teff}{T_\text{eff}}
\newcommand{\Bpole}{B_\text{pole}}
\newcommand{\Beff}{B_\text{eff}}
\newcommand{\scc}{s_\text{c}}
\newcommand{\shh}{s_\text{h}}
\newcommand{\pcc}{p_\text{c}}
\newcommand{\phh}{p_\text{h}}

\newcommand{\Reff}{R_\mathrm{eff}^\infty}

\newcommand{\msun}{{M}_\odot}
\newcommand{\xmm}{\textit{XMM-Newton}}
\newcommand{\chan}{\textit{Chandra}}

\Title{A Simple Model of Radiation from a Magnetized Neutron Star: Accreted Matter and Polar Hotspots}

\TitleCitation{Simple model of radiation from a magnetized neutron star}

\Author{Dmitry Yakovlev 
	$^{1}$
\orcidA{0000-0001-7305-2117}}

\AuthorNames{Dmitry Yakovlev}

\AuthorCitation{Yakovlev, D.}

\address[1]{%
	$^{1}$ \quad Ioffe Institute, Politekhnicheskaya 26, St~Petersburg 194021,
	Russia; yak.astro@mail.ioffe.ru}

\corres{Correspondence: yak.astro@mail.ioffe.ru}

\abstract{A simple and well known model for thermal
	radiation spectra from a magnetized neutron star is further
	studied. The model assumes that the star 
	is internally isothermal and possesses 
	dipole magnetic field ($B \lesssim 10^{14}$ G) in the outer heat-insulating layer. The heat transport
	through this layer makes the surface temperature distribution anisotropic; 
	any local surface element is assumed to emit a blackbody (BB) radiation with a local effective temperature.
	It is shown that this thermal emission is
	nearly independent of the chemical composition of insulating envelope (at the same taken averaged
	effective surface temperature). Adding a slight extra heating
	of magnetic poles allows one to be qualitatively consistent with observations
	of some isolated neutron stars.}


\begin{document}

%
\section{Yury N. Gnedin}
\label{s:yng}

This paper is dedicated to the memory of Yury N.\ Gnedin
(1935--2018). He was my senior colleague at Theoretical Astrophysics Department of Ioffe Insitute from 1971 to 1984, and we often met later, when he moved to Pulkovo Observatory. He was always interested in many scientific fields. For instance, his pet subjects were radiation transfer, neutron stars and magnetic fields (although he never lost scientific interest in a great amount of other things). When neutron stars were discovered in 1967, he was extremely enthusiastic about them and initiated ultrafast spreading of interest to these objects among colleagues. He was the first author of the first publication \cite{gnedin1969} on neutron stars at the Ioffe Institute after the discovery of these stars. He wrote -- with different co-authors -- plenty of (now) classical papers 
on neutron star physics. In particular, they include basic
formulation of the radiation transfer problem in a strongly
magnetized plasma \cite{gnedin1973}, theoretical prediction of electron cyclotron lines in the spectra of
radiation from magnetized neutron stars
\cite{gnedsun74}, the first realistic studies of ionization in the magnetized 
hydrogen atmospheres of neutron stars \cite{GPT74}. Also, he was greatly fond of
history.

\begin{figure*}
	\centering
	\includegraphics[width=0.95\textwidth]{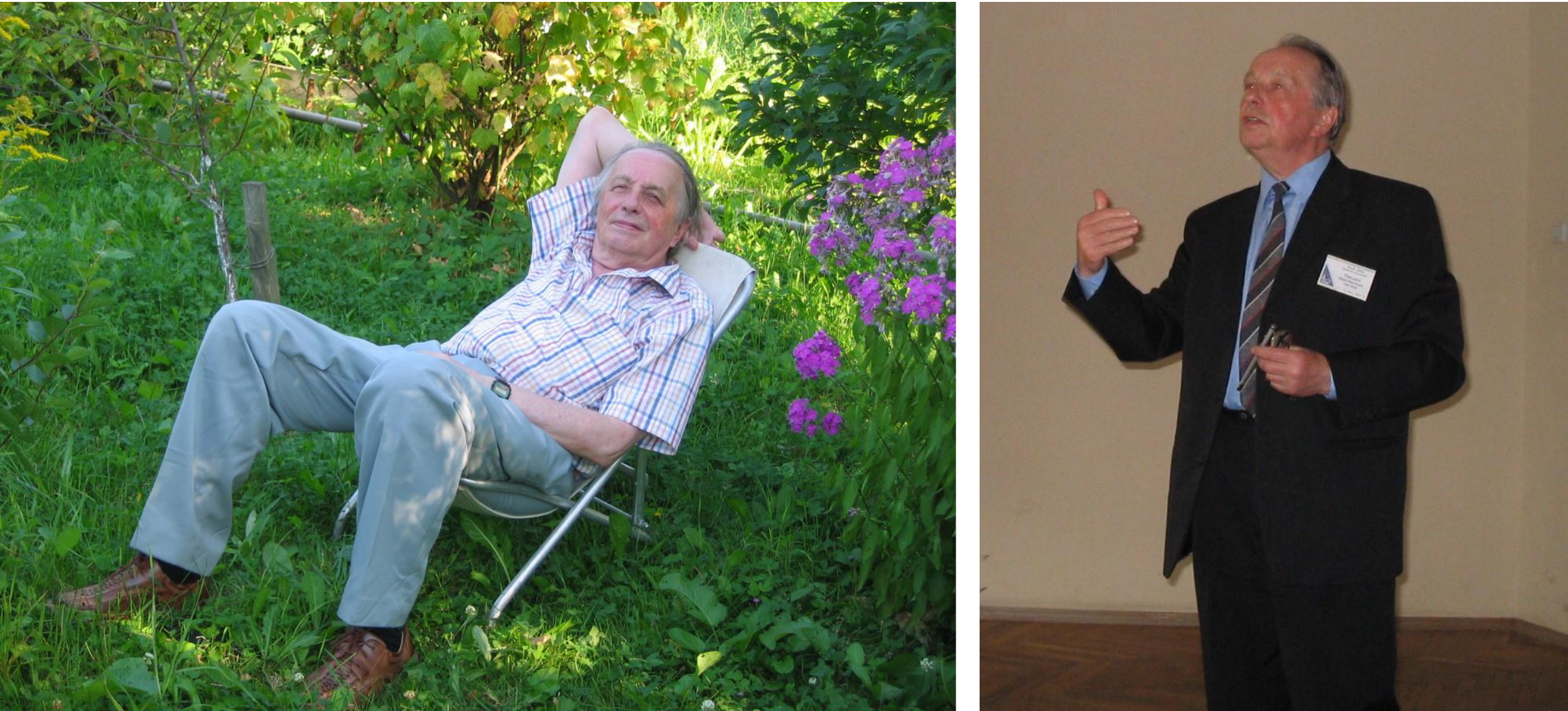}%
	\caption{
		Yury N. Gnedin. {\it Left:} A rare moment of
		leisure on his birthday, August 13, 2007, near his summer cottage. {\it Right:} At a conference (June 2004). Courtesy 
		to Oleg Gnedin.	
	}
	\label{f:young}
\end{figure*}

Although we did not work much together I, as many others, respected Yury Gnedin very much. When he could
help, he really did, and he was a Real Gentleman -- friendly and
open to any one, from a prominent scientist to a hopeless student.  I will
always remember the wonderful atmosphere, created by Yury Gnedin, and warm
hospitality of his family and home. It is amazing that both his sons have grown up into excellent scientists.

I present a small piece of neutron star physics on thermal radiation 
from magnetized neutron stars. Hopefully, he would like it.

\section{Introduction}
\label{s:introduc}

I will mostly consider thermal X-ray radiation from the
surface of a middle-aged ($t \lesssim 10^6$ yr) isolated
magnetized neutron star. The radiation is supposed to
emerge from warm neutron star interior and emitted from
a thin atmosphere or condensed surface 
(e.g. \cite{Potekhin14}), where the radiation
spectrum is formed. The interiors of the star are thought to
be nearly isothermal because of high thermal conductivity of dense
matter. Nevertheless, the interiors are thermally
insulated from the surface by a thin heat blanketing envelope,
where the thermal conduction is poorer and sensitive to the
chemical composition and
magnetic field. The field redirects the emerging heat flow and results
in anisotropic surface temperature distribution.
The thermal surface emission can also become anisotropic.

The anisotropy of observed surface radiation is
used to infer magnetic field strength and geometry, 
composition of the surface layers,
global parameters of the star (such as mass and radius),
as well as important parameters of superdense matter in
stellar interiors
(see, e.g., \cite{Potekhin14,PPP2015a} and references therein).

Here I study a simple model of thermal emission 
from magnetized neutron stars with isothermal interiors, assuming
the star emits a blackbody (BB) radiation with local effective
surface temperature $\Ts$, that varies over stellar surface under the
effects of dipole surface magnetic field. This model has been
invented long ago by \citet{Greenstein}, elaborated in the literature 
(e.g. \cite{page95,page96,PY01,PYCG03,geppert2006,zane-turolla2006})
and reviewed in \cite{PPP2015a,BPY21}. Here I point out some
properties of the surface emission which, to the best of my knowledge,
have not been studied in the literature. 

Logically, this paper continues the previous one
\cite{yak21} (hereafter Paper I) which shows that simple thermal spectra
of magnetized neutron stars can be accurately approximated
by two-BB (2BB) models. Sect.\ \ref{s:model} 
outlines  
theoretical formalism. Sect.\  \ref{s:acc} is devoted
to the effects of chemical composition
of the blanketing envelopes and Sect.\ \ref{s:hotspots}
extends the solution to the case, in which
magnetic poles contain additional hot spots.
The conclusions are formulated in Sect.\ \ref{s:discuss}.

It is important to mention more complicated 
neutron star emission models. They include sophisticated
magnetic field effects on thermal emission, leading to specific 
spectral, polarization and angular properties of radiation 
(see, e.g. \cite{2002ZAVPAV,PPP2015a,gonzalezea2016,grandis21}),
which are not reproduced by the given model.

\section{Simple model spectra}
\label{s:model}

\subsection{General formalism}
\label{s:general formalism}

Calculation of radiation spectral flux 
from a spherical neutron star under formulated assumptions
is simple. Here is the summary using the notations of Paper I.
Any small surface temperature element
emits like a BB, and the flux is obtained by intergating
over a vizible part of the surface, taking intro account
that emitted quanta propagate in Schwarzschild space-time
and demonstrate gravitational redshift 
of photon energies and light bending.
The effects of General Relativity are specified
by the compactness parameter $\xg=\rg / R$,
where  $\rg=2GM/c^2$ is the Schwartschield radius,
$M$ is the gravitational star's mass, $R$ is
the circumferential radius, $G$ is the gravitational
constant and $c$ is the velocity of light. One deals either
with local quantities at the stellar surface (e.g., the
local surface temperature $\Ts$) or with 
the quantities detected by
a distant observer. The letter ones will be often
marked by the symbol `$\infty$'. 
For instance, $\Ts^{\infty}=\Ts \, \sqrt{1-\xg}$ 
is the redshifted surface temperature. As an
exclusion, we will denote local (non-redshifted) photon energy
by $E_0$, and the redshfied energy by $E  \equiv E_{\infty} =
E_0 \sqrt{1-\xg}$.

Let $F_E^{\infty}$ be a radiative spectral flux density
[erg cm$^{-2}$ s$^{-1}$ keV$^{-1}$] detected 
at a distance $D \gg R$. It is customary to express $F_E^{\infty}$ as
\begin{equation}
F_E^{\infty}=\frac{R^2}{D^2}\,H_E^{\infty}, 
\label{e:H}   	
\end{equation}  
where $H_E^{\infty}$ is the effective flux, that
is formally independent of $D$. It can be 
calculated as an integral of the radiating 
surface flux over the visible part of the
surface,
\begin{equation}
H_E^{\infty}=\frac{15 \SB }{16 \pi^5 k_{\rm B}^4}\, 
\int_{\rm viz} \dd \Omega_{\rm s}\,
\frac{(1-\xg)^{-1}\, {\cal P}  E^3}{\exp(E/\kB \Ts^{\infty})-1},
\label{e:H(Einfty)} 
\end{equation} 
where $\sigma_{\rm SB}$ is the
Stefan-Boltzmann constant, $\kB$ is the Boltzmann constant, 
${\rm d}\Omega_{\rm s}$ is a surface solid angle element, and
${\cal P}$ is the light bending function
(e.g. \cite{2002Beloborodov}, 
\cite{2003Poutanen}, \cite{Potekhin14} and \cite{2020Poutanen}).

It is convenient to integrate over the star's surface and
calculate the bolometric luminosity
of the star, $L_\text{s}$, as well as the averaged non-redshifted
effective surface temperature $\Teff$  (e.g. \cite{PY01}),
\begin{equation}
	L_\text{s}=\SB R^2 \int_{4 \pi} \dd \Omegas \, \Ts^4
	\equiv 4 \pi \SB R^2 \Teff^4.
	\label{e:Ls}
\end{equation}

For a uniform surface temperature ($\Ts=\Teff$), 
one immediately gets the standard BB flux,
\begin{equation}
H_E^{\rm BB \infty}=\frac{15 \SB }{4 \pi^4 k_{\rm B}^4}\,
\frac{(1-\xg)^{-1}\, E^3}{\exp(E/\kB \Teff^\infty)-1},
\label{e:BB} 
\end{equation} 
and the bolometric effective flux 
\begin{equation}
H_{\rm bol}^{\rm BB\infty}=\int_0^\infty H_E^{\rm BB \infty}\,
\dd E=\frac{\SB \Teff^{\infty 4}}{1-\xg}.
\label{e:BBbol}
\end{equation}

\subsection{Input parameters}
\label{s:input}

Eq.\ (\ref{e:H(Einfty)}) allows one to compute
thermal spectra for any given temperature
distribution $\Ts$ over neutron star surface.
We focus on the distribution created by
a dipole magnetic field (with the field strength
$\Bpole$ at magnetic poles) due to anisotropic
heat transport in a thin (maximum a few hundreds
meters) heat blanketing envelope. The 
input parameters are $M$, $R$, chemical composition
of the blanketing envelope, and a nonredshifted
temperature $\Tb$ at its bottom (at density
$\rhob \sim 10^{10}$ $\gcc$; see e.g. \cite{BPY21}).
 The local $\Ts$ 
is usually determined by solving local quasistationary 
radial heat transport within the envelope mediated by an effective
radial thermal conductivity. For studying the
thermal surface emission it is profitable to use
$\Teff$ [see Eq.\ (\ref{e:Ls})] instead of $\Tb$.

\subsection{2BB representation}
\label{s:2BBimage}

According to Paper I, the spectral fluxes
$H_E^\infty$,  computed from Eq.\ (\ref{e:H(Einfty)})
for iron heat blankets,
are accurately fitted by a familiar 2BB model, 
\begin{equation}
	H_E^\infty=\scc H_E^{\rm BB \infty}(T_{\rm eff c})+
	\shh H_E^{\rm BB \infty}(T_{\rm eff h}).
	\label{e:2BBfit}    
\end{equation} 
Here $H_E^{\rm BB \infty}(\Teff)$ is given by Eq.\
(\ref{e:BB}); `c' and `h' refer, respectively, to colder and hotter
BB components. Any
fit contains four parameters, which
are two effective temperatures $T_{\rm eff c}$ and $T_{\rm eff h}$, 
and two fractions of effective radiating surface areas, $\scc$ and 
$\shh$. Instead of $T_{\rm eff c}$ 
and $T_{\rm eff h}$, it is convenient to introduce two 
dimensionless parameters
$\pcc=T_{\rm effc}/\Teff$ and $\phh=T_{\rm effh}/\Teff$,
with $\pcc < \phh$. In Paper I the fits have been done 
for a number of representative values of $M$, $R$, 
$\log \Teff$[K] (from 5.5 to 6.8), $\log \Bpole$[G] (from
11 to 14), photon energies ($0.064<E \lesssim 40$ keV,
removing those $E$ at which the fluxes are
negligibly small) and inclination angles $i$ (between line of sight and the
magnetic axis). Typical relative fit errors
have not exceeded a few per cent, meaning the fits
were really good, 
providing excellent analytic representation
of original computed data. Therefore, thermal X-ray
spectral fluxes of magnetized neutron stars within the given
model are nearly identical to 2BB spectral models. 

Moreover, as shown in Paper I, it is sufficient to calculate
the fluxes $H_E^\infty$ for two observation
directions which are (i) pole observation,
$i=0$, to be denoted as $H^{\parallel \infty}_E$,
and (ii) equator observations, $i=90^\circ$, to be
denoted as $H_E^{\perp \infty}$. If these 
fluxes are known,  
the radiation flux in any direction $i$ is accurately
approximated as
\begin{equation}
	H^{i \infty}_E=H_E^{\parallel \infty} \cos^2 i + H_E^{\perp \infty} \sin^2 i.
	\label{e:Hanyi}
\end{equation}
We will see (Sect.\ \ref{s:hotspots}) that that this approximation is more restrictive than Eq.\ (\ref{e:2BBfit}).

This accurate representation of numerically 
calculated $H^{i \infty}_E$ by a
map of four fit parameters 
($\pcc$, $\phh$, $\scc$, $\shh$) in
space of input parameters has to be taken with
the grain of salt. The problem is that the fluxes
$H^{i \infty}_E$ are close to 1BB fluxes (\ref{e:BB}),
which leads to some degeneracy of fit parameters (with
respect to a fit procedure and a choice of grid points). 
On the other hand, the assumed surface temperature
distribution model and the local BB emission model
are definitely approximate by themselves, so that
too accurate fitting of $H^{i \infty}_E$ is actually
purely academic. However, 
these results can be helpful for interpretation of
observations, especially because the 2BB model is often
used by observers.

\section{Iron and accreted heat blankets}
\label{s:acc}

Paper I considered heat blankets made of iron. Here we employ
a more general model 
\cite{PCY97,PY01,PYCG03}, in which
a heat blanket consists of shells 
(from top to bottom) of hydrogen, helium, carbon and
iron. Light elements (H, He, C) are thought to appear because of accretion of H and/or He and further burning
into C. Iron is either formed at neutron star birth
or is a final result of carbon burning. 
The mass of lighter elements
$\Delta M$ in the heat blanket is treated as a free parameter.
The heat blanketing envelope may fully consist of 
iron (Paper I) or contain some mass $\Delta M
\lesssim 10^{-6} \, \msun$ of lighter elements.
Lighter elements affect  
the surface temperature distribution and thermal surface 
spectral fluxes. 

To simplify the task, we will study 
fully accreted blankets and conclude on the partly
accreted ones in the end of Sect.\ \ref{s:fluxes}.

\subsection{$\Ts$ distributions}
\label{s:Palex}

\begin{figure*}
	\centering
	\includegraphics[width=0.33\textwidth]{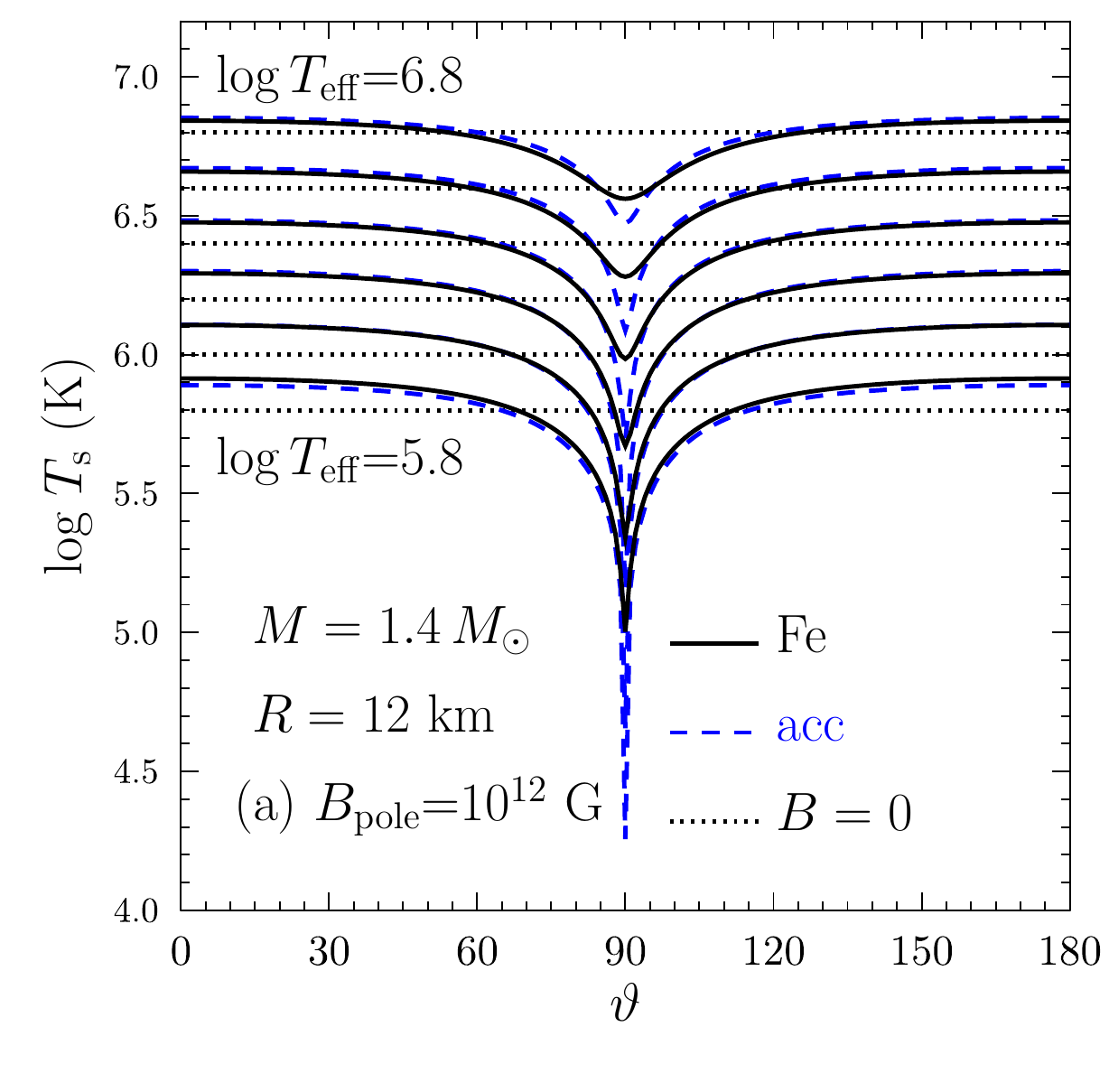}%
	\hspace{-5mm}
	\includegraphics[width=0.33\textwidth]{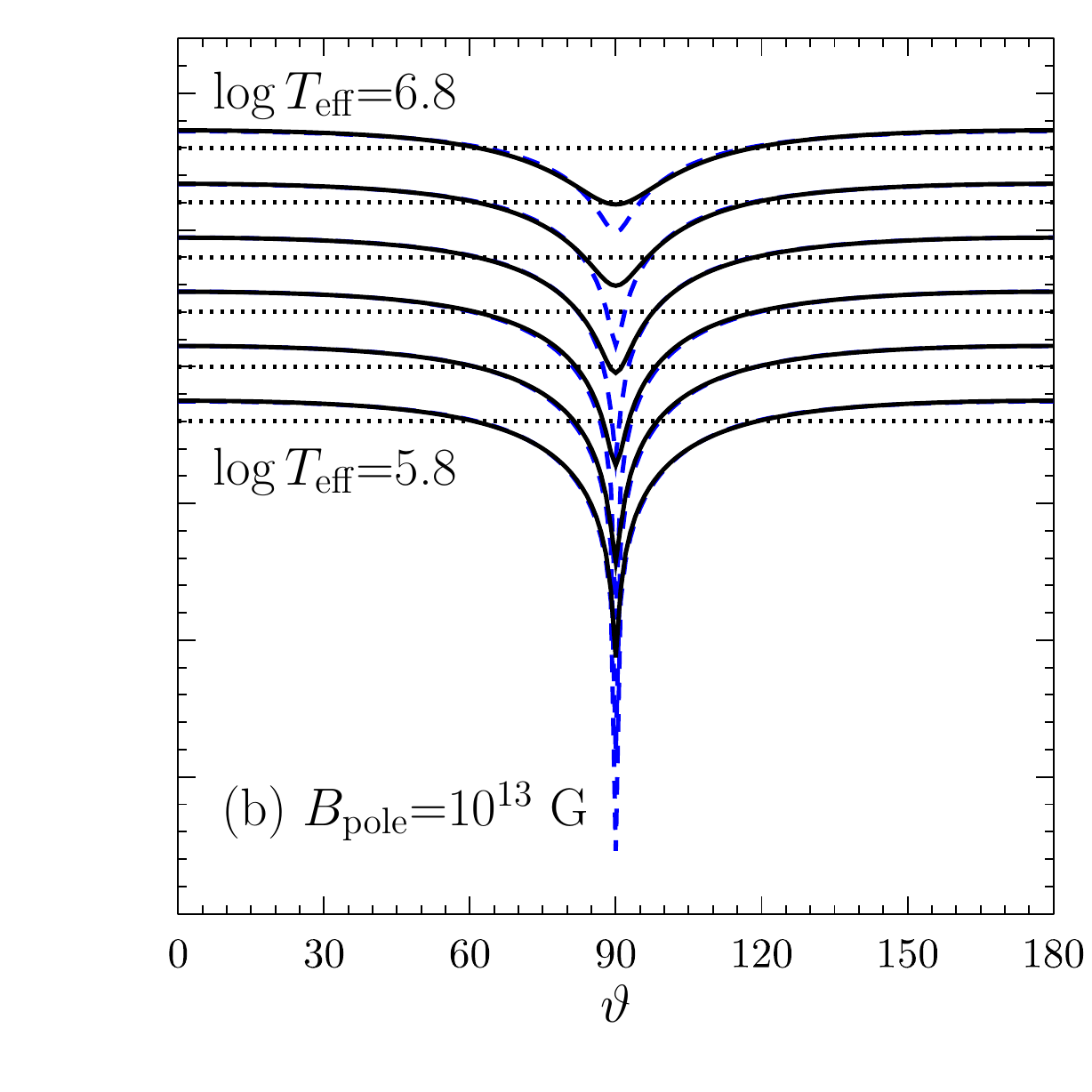}%
	\hspace{-5mm}
	\includegraphics[width=0.33\textwidth]{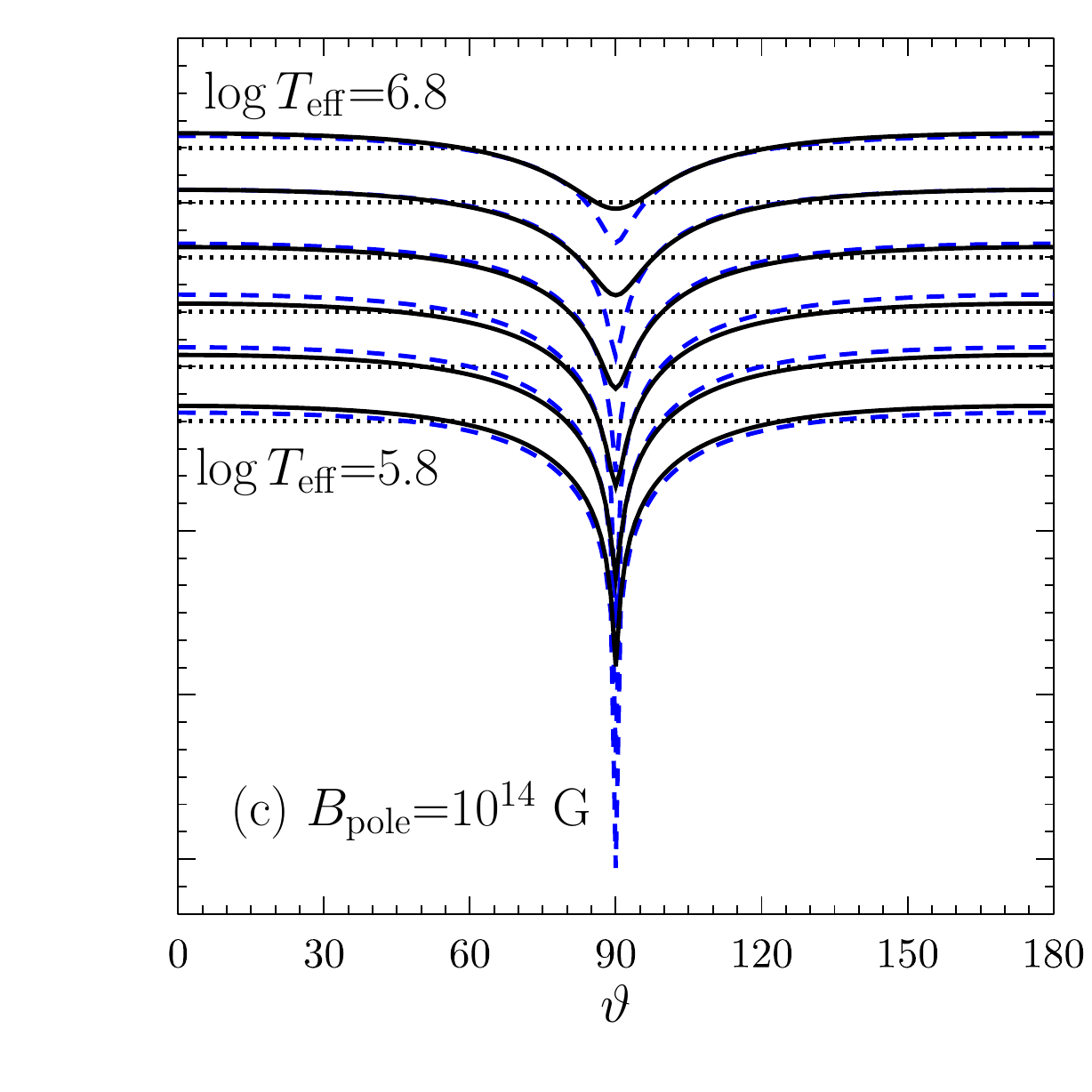}%
	\caption{
		Effective surface temperature $\Ts$ versus colatitude
		$\vartheta$, from magnetic pole ($\vartheta=0$) to equator 
		($\vartheta=90^\circ$) 
		and the opposite magnetic pole  ($\vartheta=180^\circ$)
		of a neutron star with 
		$M=1.4\,\msun$ and $R=12$ km at 
		three values of $\Bpole=10^{12}$, $10^{13}$ and
		$10^{14}$~G [panels (a--c)]. 
		The curves are plotted for six effective surface temperatures, 
		$\log \Teff$ [K]=5.8, 6.0 \ldots 6.8. Solid curves 
		refer to the heat-blanket model made of iron,
		dashed cures refer to the heat blanket of fully
		accreted matter; dotted curves correspond to
		the field-free star.			
	}
	\label{f:Ts}
\end{figure*}

Our surface temperature distribution $\Ts$  
is axially symmetric and symmetric with respect
to the magnetic equator. 
For illustration, Fig.\ \ref{f:Ts} presents 
calculated $\Ts$ for a star with
$M=1.4 \, \msun$ and $R=12$ km 
[$\xg$=0.344 and the surface gravity
$g_{\rm s}=GM/(R^2 \sqrt{1-\xg})=1.59 \times 10^{14}$ cm~s$^{-2}$]. 
The effective surface
temperatures are shown versus colatitude $\vartheta$,
which varies from $\vartheta=0$ at the north pole,
to $\vartheta=90^\circ$ at the magnetic equator, and 
then to $\vartheta=180^\circ$ at the south pole.
Three panels (a-c) correspond
to three values of the magnetic field
$\Bpole=10^{12}$, $10^{13}$ and $10^{14}$~G
at the pole. 
Each panel displays $\Ts(\vartheta)$ for six values
of $\log \Teff$ [K]= 5.8, 6.0, \ldots 6.8. Solid lines
are plotted for the heat blankets made of iron, while dashed
lines are for the fully accreted blankets. 
The iron and accretted matter have different
thermal conductivities and, hence, different $\Ts(\vartheta)$
distributions at the same $\Teff$.  The
dotted lines correspond to
a non-magnetic star to guide
the eye (they give $\Ts=\Teff$).

Heat conduction is mainly radiative in the outer non-degenerate layer.
In deeper layers, where the electrons become degenerate,
heat is mostly transported by electrons. The field
affects also thermodynamic properties of the matter, for
instance, the pressure (e.g. \cite{HPY07}). 
The magnetic field effects, which
regulate the anisotropy of the surface temperature
distribution, are twofold. 

The effects of the first type
are the classical effects of electron rotation about magnetic field lines. 
They are especially important near the 
equator, where the heat, emergent from the stellar
interiors, propagates mainly across field lines
and is thus suppressed. This may strongly
enhance thermal insulation of the 
equatorial part of the heat blanketing
envelope, producing rather
narrow equatorial dips of $\Ts(\vartheta)$
(the cold equatorial belt, Fig.\ \ref{f:Ts}). 
The dips for the accreted matter are seen to be much stronger
than for the iron matter, especially at lower $\Teff$.

The effects of second type 
are associated with
Landau quantization of electron motion across 
$\bm{B}$-lines. They become more efficient at higher $B$ near
magnetic poles, where they make the heat insulating layers
more heat transparent and tend to increase $\Ts$. These
relatively warmer polar `caps' are wide
(Fig.\ \ref{f:Ts}); $\Ts$ increases inside them, if we
approach a pole, but not dramatically.  

As long as $\Bpole$ is 
rather weak ($\Bpole \lesssim 10^{11}$ G), the classical
effects stay more important, producing a pronounced
colder equatorial belt, but weakly affecting the polar
zones. The surface becomes overall colder,
compared with the field-free star with the same internal
temperature $\Tb$ (e.g., see Figs.\ 22 and 23 in
\cite{BPY21}).
With increasing $\Bpole$, the quantum effects become 
dominant, creating hot polar zones
and making the surface overall 
hotter (at the same $\Tb$). At
$\Bpole \gtrsim 10^{12}$ G, the surface becomes hotter than at $\Bpole=0$;
hot polar zones make the equatorial belt 
less significant.

It is remarkable (Fig.\ \ref{f:Ts}) that 
outside the equatorial belt, at fixed $\Teff$,
the surface temperature profiles $\Ts(\vartheta)$ for iron
and accreted heat blankets are mainly close. 
Moreover, these profiles
almost coincide at $\Bpole \approx 10^{13}$~G. The latter
conclusion is independent of specific values of $M$ and $R$.

\subsection{Spectral fluxes}
\label{s:fluxes}

\begin{figure*}[t]
	\centering
	\includegraphics[width=0.33\textwidth]{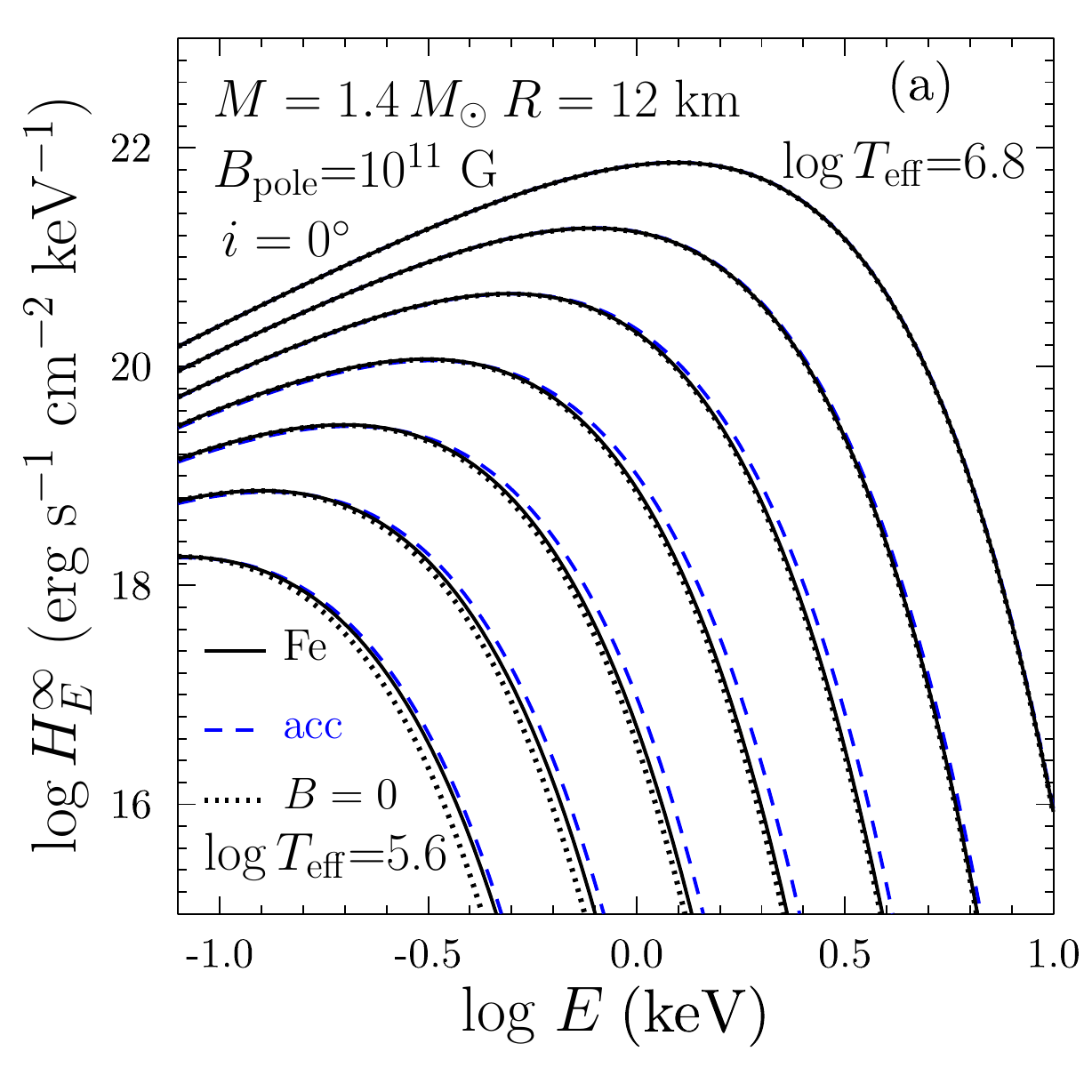}%
	\hspace{-5mm}
	\includegraphics[width=0.33\textwidth]{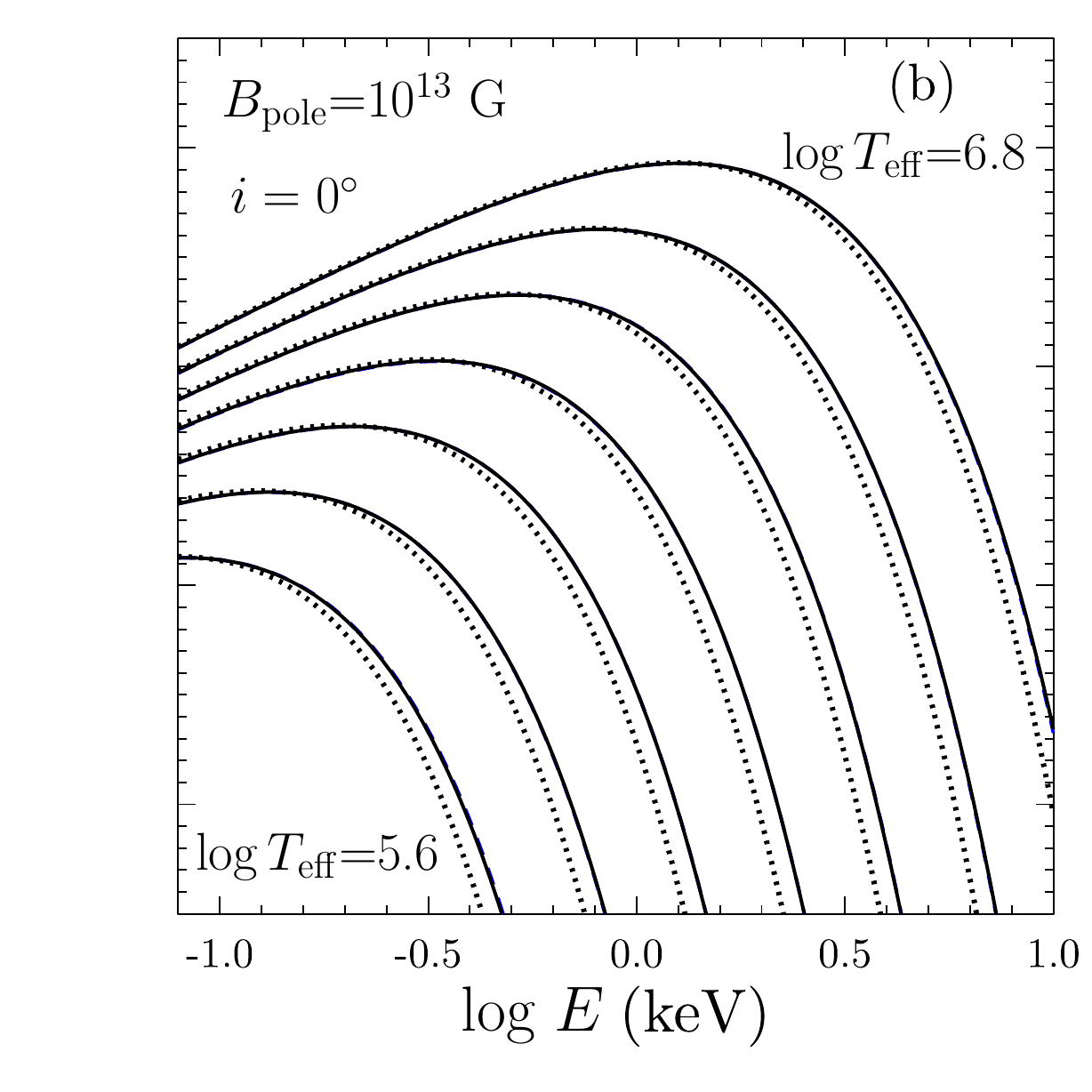}%
	\hspace{-5mm}
	\includegraphics[width=0.33\textwidth]{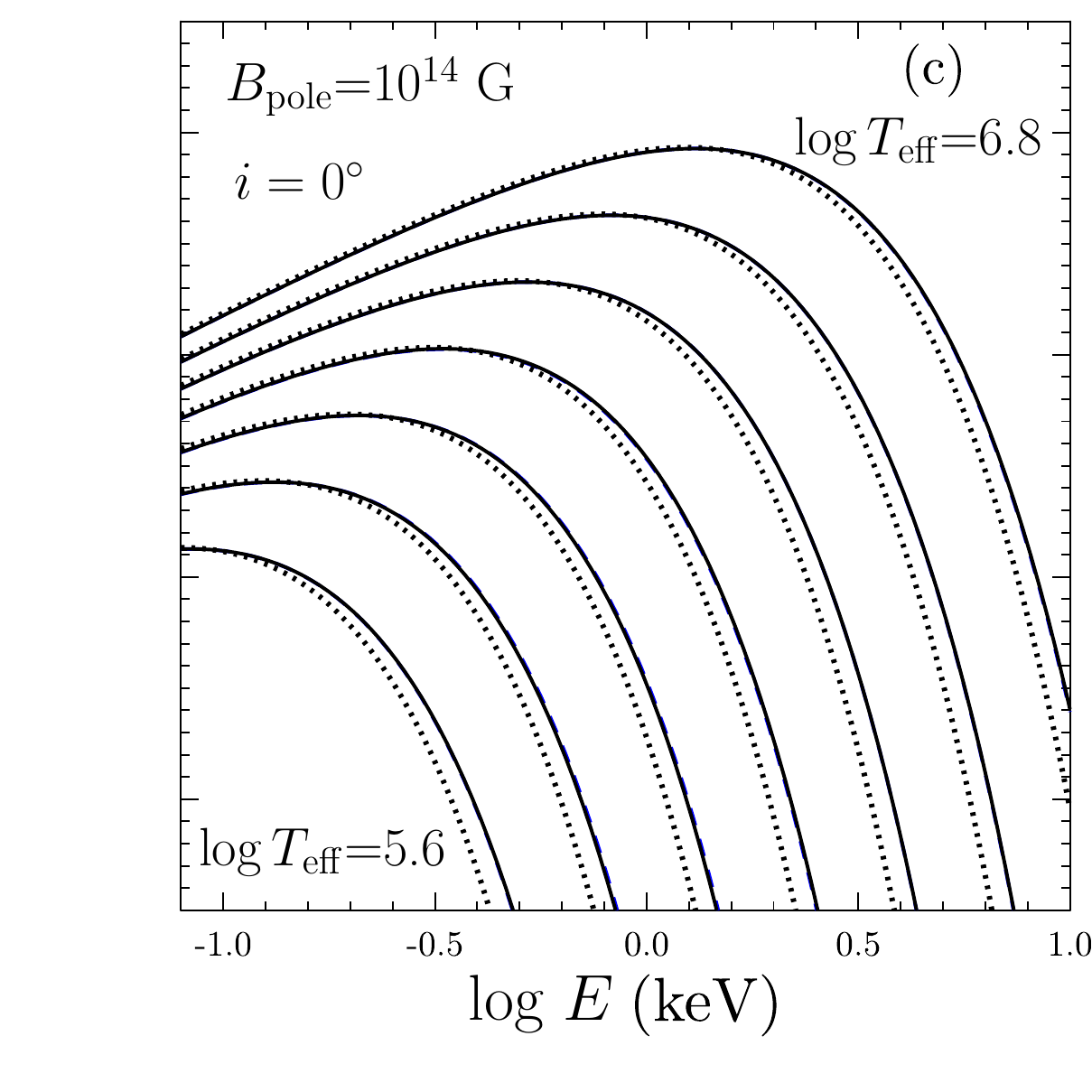}%
	\caption{
		Thermal spectral fluxes observed from magnetic poles 
		for a neutron star with 
		$M=1.4\,\msun$ and $R=12$ km at 
		three values of $\Bpole=10^{11}$~G (a), $10^{13}$~G (b),
		and $10^{14}$~G (c) at $\log \Teff=$
		5.6, 5.8,\ldots,6.6 assuming iron (solid lines) or
		accreted (dashed lines) heat blankets. The dotted lines
		refer to non-magnetic star to guide the eye. 			
	}
	\label{f:Hparallel}
\end{figure*}

Despite of dramatic interplay of the opposite effects and
substantial temperature anisotropy, the surface-integrated
spectral fluxes $H_E^\infty$ of thermal radiation behave 
smoothly under variations of $\Bpole$ and/or
$\Teff$. This is because of strong effects 
of General Relativity, which
smooth out photon propagation from the surface to the
distant observer \cite{page95,PY01,PYCG03}. As noted
in the cited papers, General
Relativity `hides' magnetic effects on thermal
surface emission. Very large magnetic fields, with
$\Bpole \gtrsim 10^{14}$ (typical of magnetars),
start to affect the surface emission much stronger but
such fields are not considered here. They deserve a
special study. In particular, their heat-insulating
envelopes can become thick and the 1D approach to calculate
the heat insulation within them can be questionable.

We have checked that the spectral fluxes,
calculated for the accreted envelopes, possess
the same properties 
(Sect.\ \ref{s:2BBimage})
as for the iron envelopes. In particular, the
flux remains to be close to that calculated in the
1BB approximation at uniform 
surface temperature $\Ts=\Teff$ (see, e.g., Fig.\ 1 
in Paper I). It is disappointing,
meaning that such a flux, taken at $\Teff$ as an input parameter,
carries little information on the magnetic field.
Then one should study deviations from
the 1BB approximation. It is important that
if one fixes $\Tb$ and varies $\Bpole$, one obtains
noticeable $\Ts$ and flux variations (as demonstrated, for
instance, in Figs. 23 and 24 of \cite{BPY21}) but these
variations are accompanied by variations of $\Teff$. 
This would be a good theoretical construction, 
if one knew $\Tb$. However, the 
observer can measure $\Teff$ and wishes to infer $\Tb$,
which is not easy because at
fixed $\Teff$ the theoretical spectral flux is 
slightly sensitive to~$\Tb$.

\begin{figure}[h]
	\includegraphics[width=0.37\textwidth]{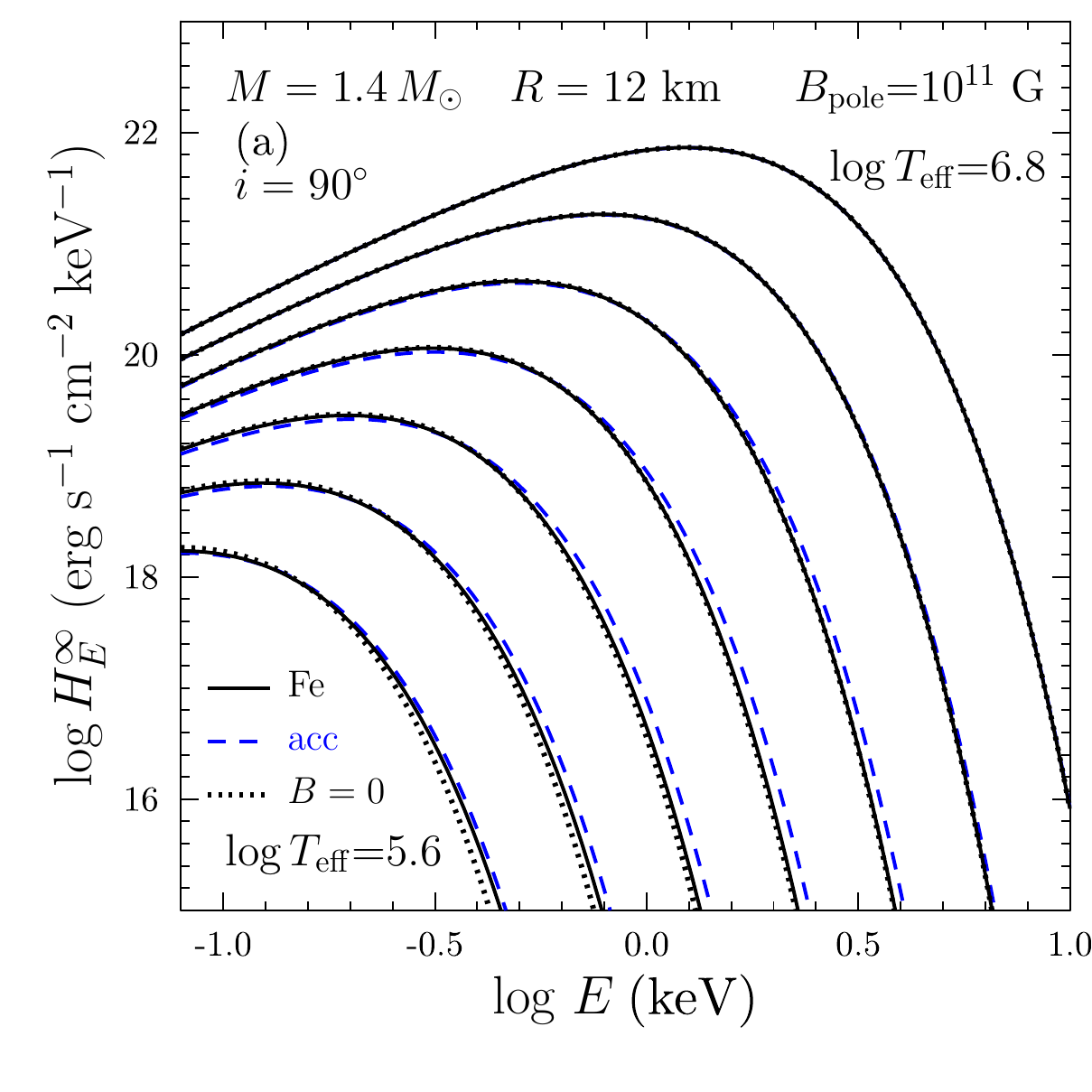}%
	\hspace{-7mm}
	\includegraphics[width=0.37\textwidth]{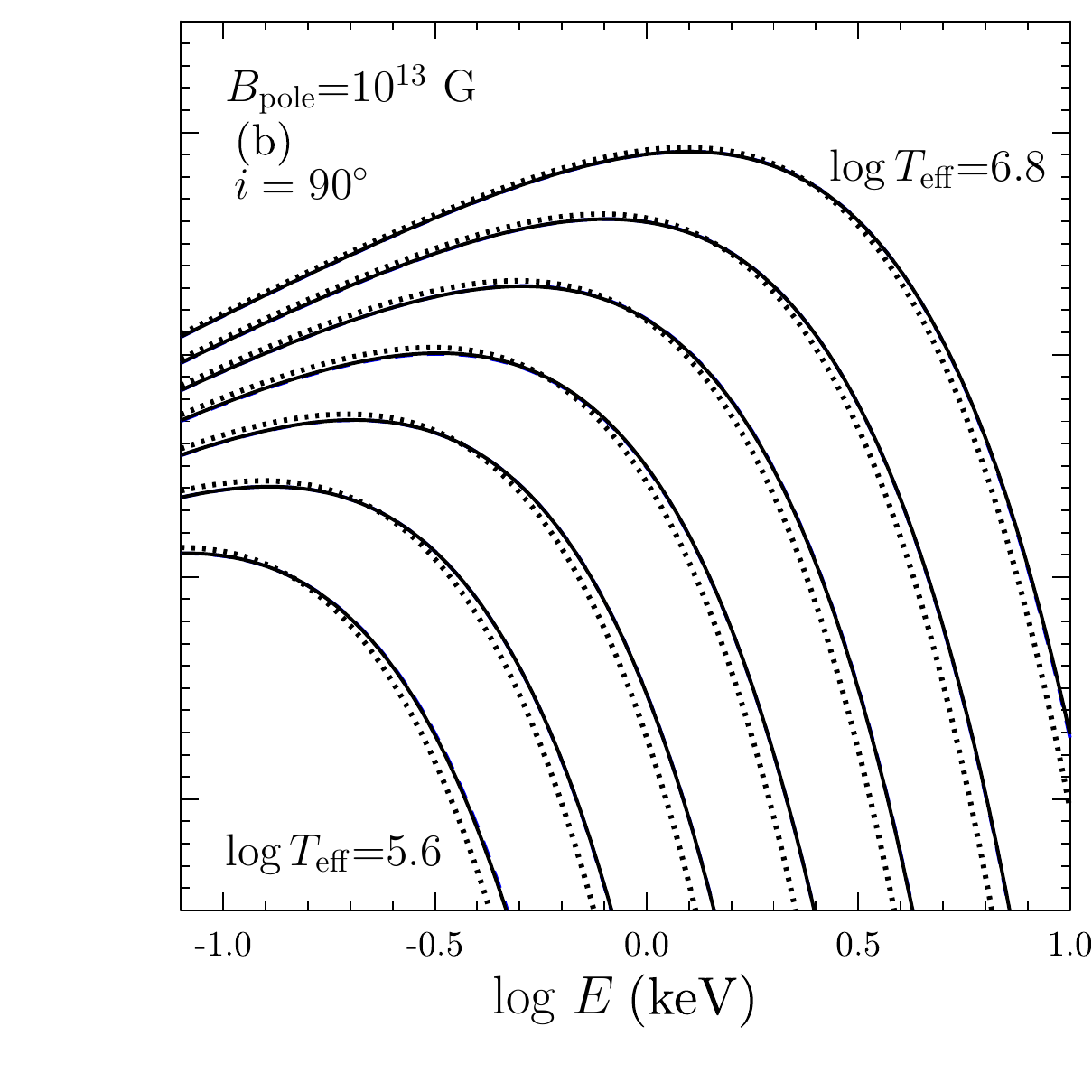}%
	\caption{
		Same as in Fig.\ \ref{f:Hparallel} but for equator
		observations at 
	    $\Bpole=10^{11}$~G (a) and 
		$10^{13}$~G (b). 			
	}
	\label{f:Hperp}
\end{figure}
 
Fig. \ref{f:Hparallel} 
shows thermal spectral flux densities $H_E^{\parallel\infty}$ for polar observations 
of the same star as in Fig.\ \ref{f:Ts} (at  seven
values of $\Teff$ and three values of
$\Bpole$). The fluxes are plotted
in logarithmic scale versus decimal logarithm of the redshifted 
photon energy $E$.  For each value of $\Teff$ we present three curves.  
The solid curves give the fluxes radiated from the star with
iron heat blanket.
The dashed curves present similar fluxes for accreted
heat blankets. 
The dotted curves demonstrate 
the 1BB model, Eq.\ (\ref{e:BB}), with given constant $\Ts=\Teff$ (as if the magnetic field is absent).

\begin{figure*}
	\includegraphics[width=0.27\textwidth]{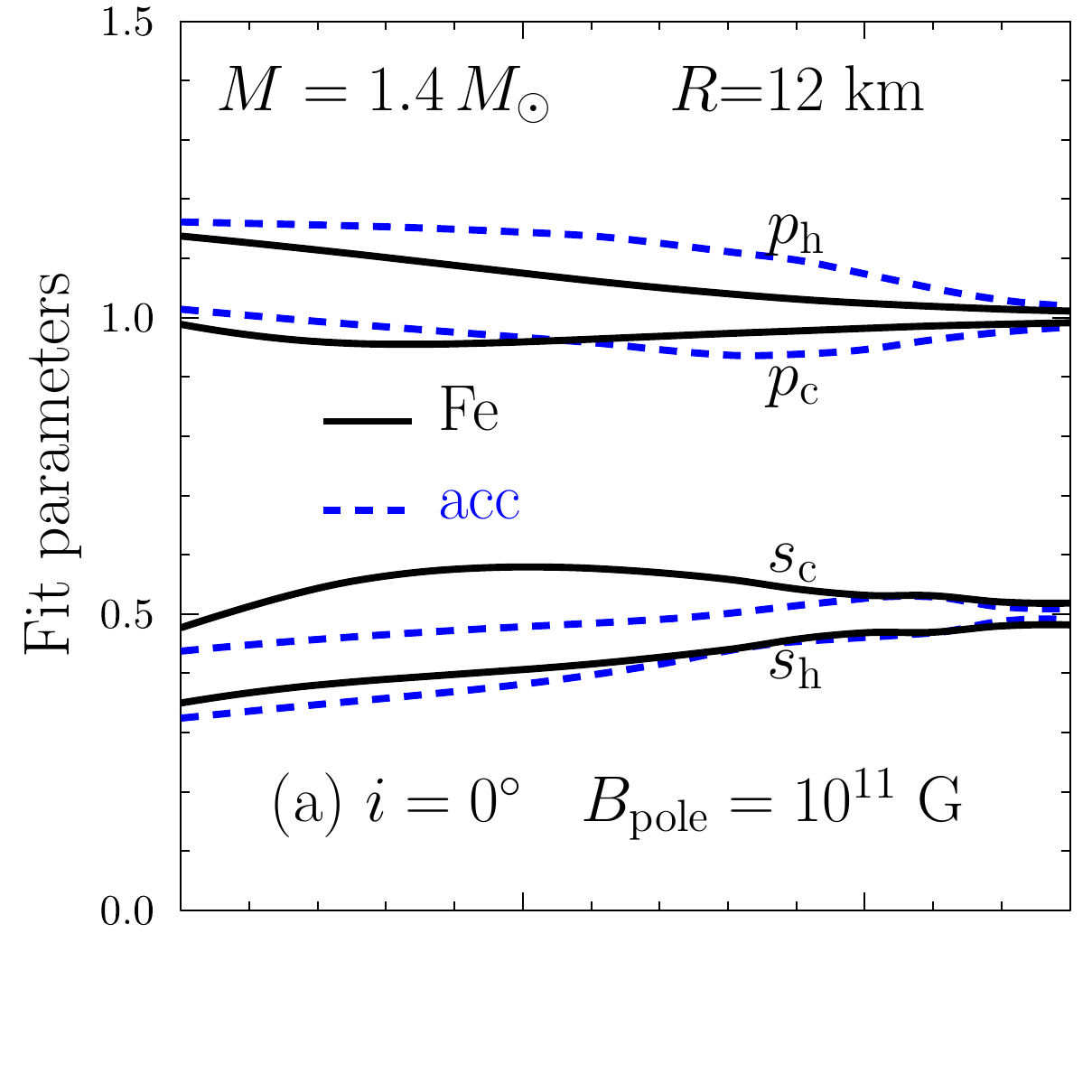}%
	\hspace{-7mm}
	\includegraphics[width=0.27\textwidth]{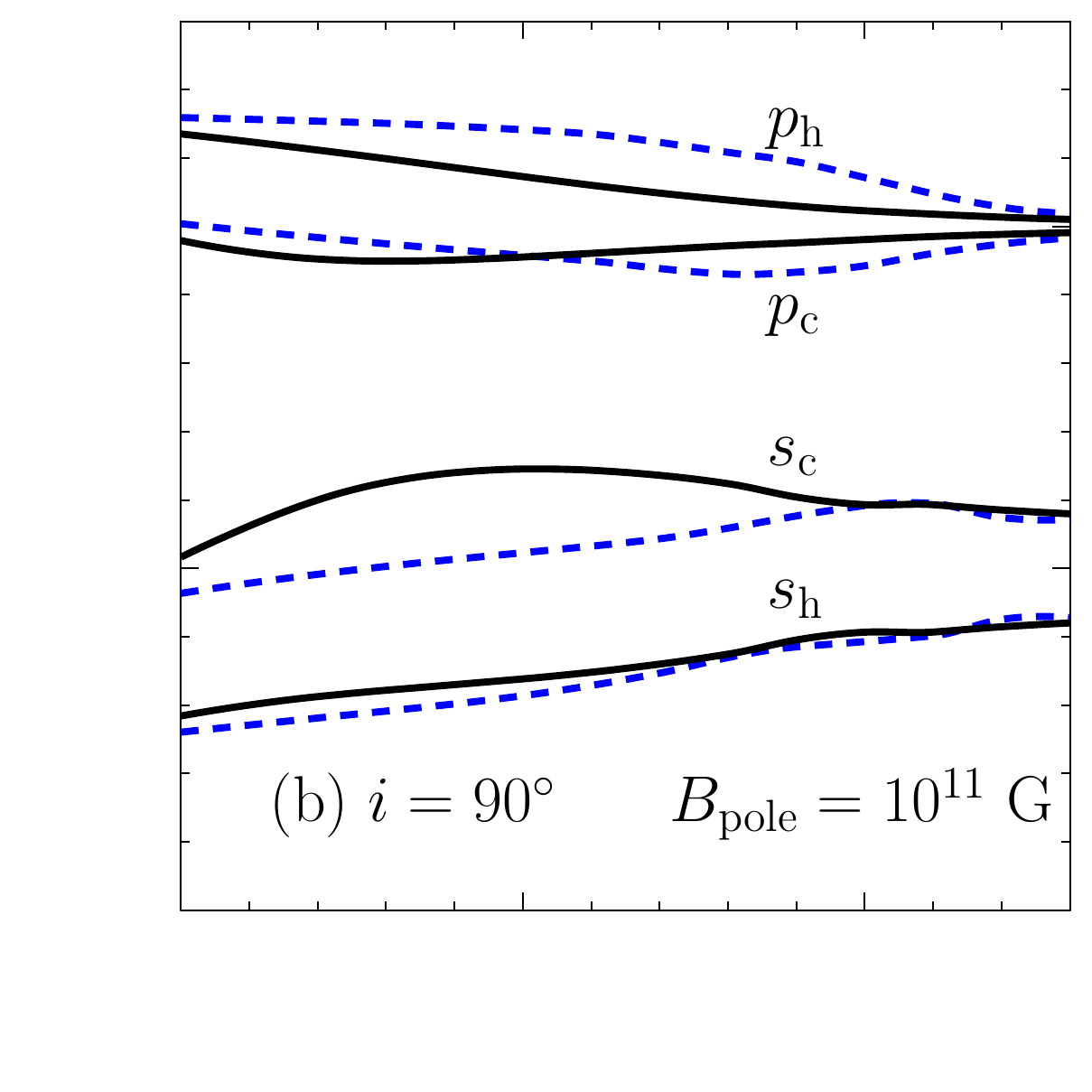}%
	\hspace{-5mm}
	\includegraphics[width=0.27\textwidth]{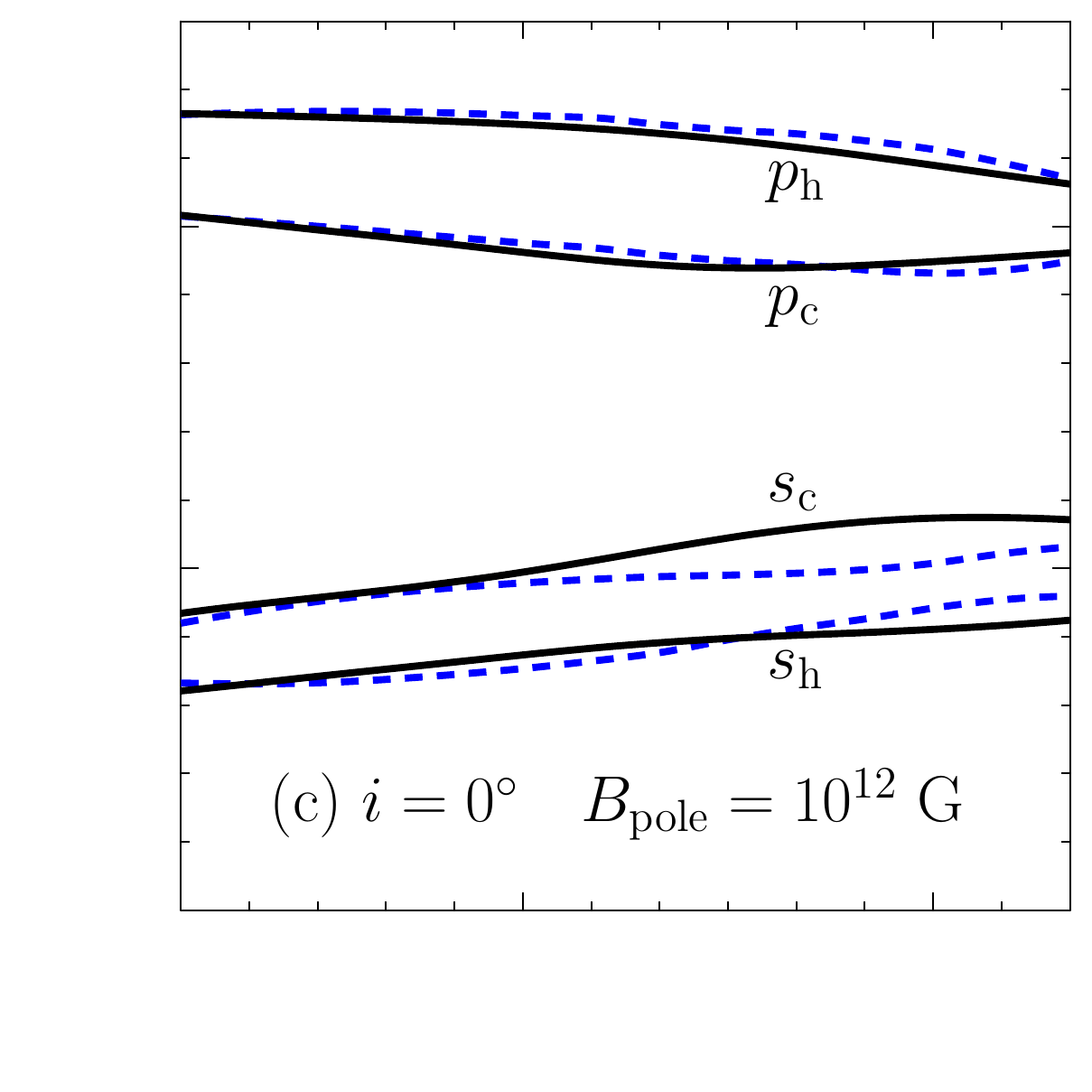}%
    \hspace{-7mm}
    \includegraphics[width=0.27\textwidth]{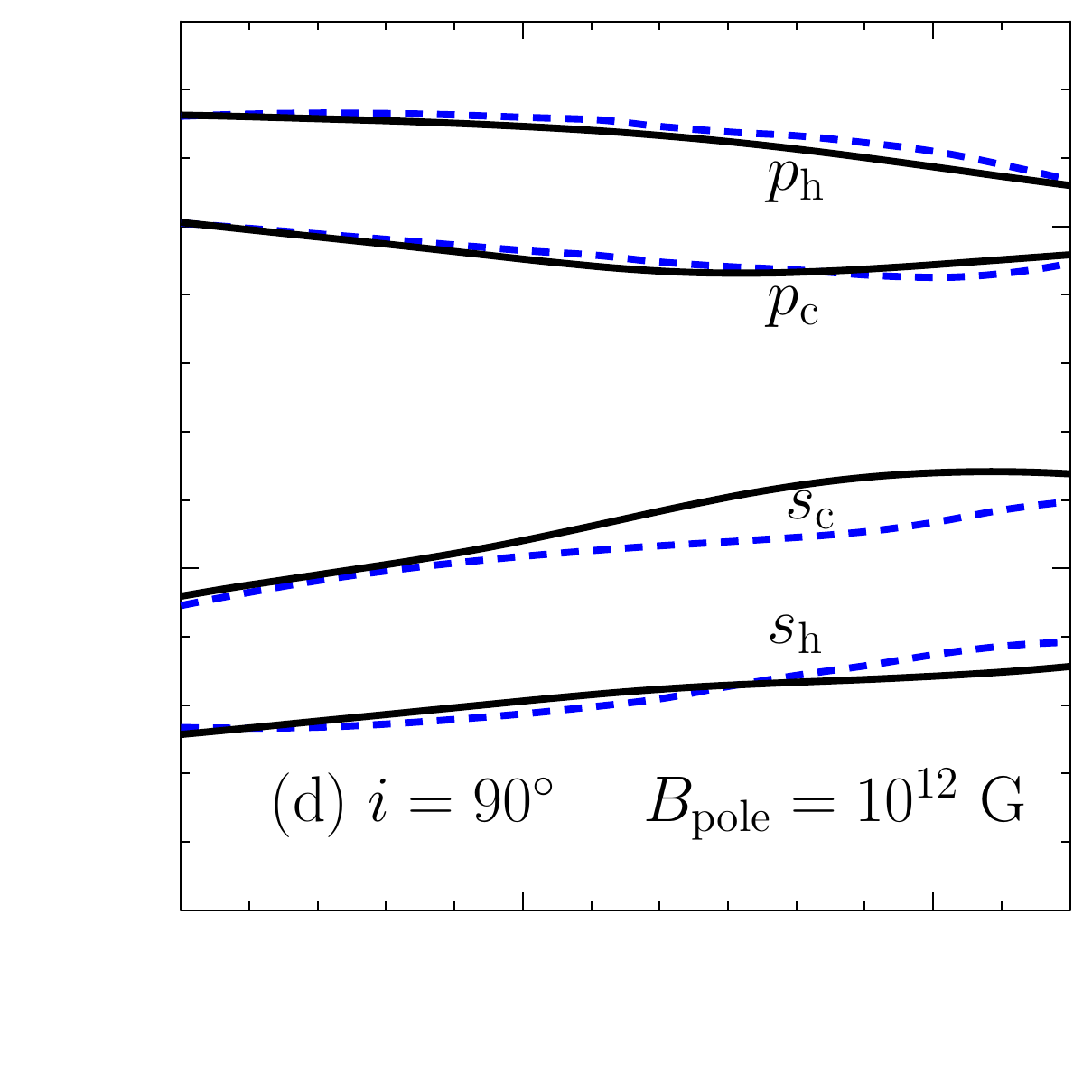}%
    \vspace{-7mm}
    \\
    \includegraphics[width=0.27\textwidth]{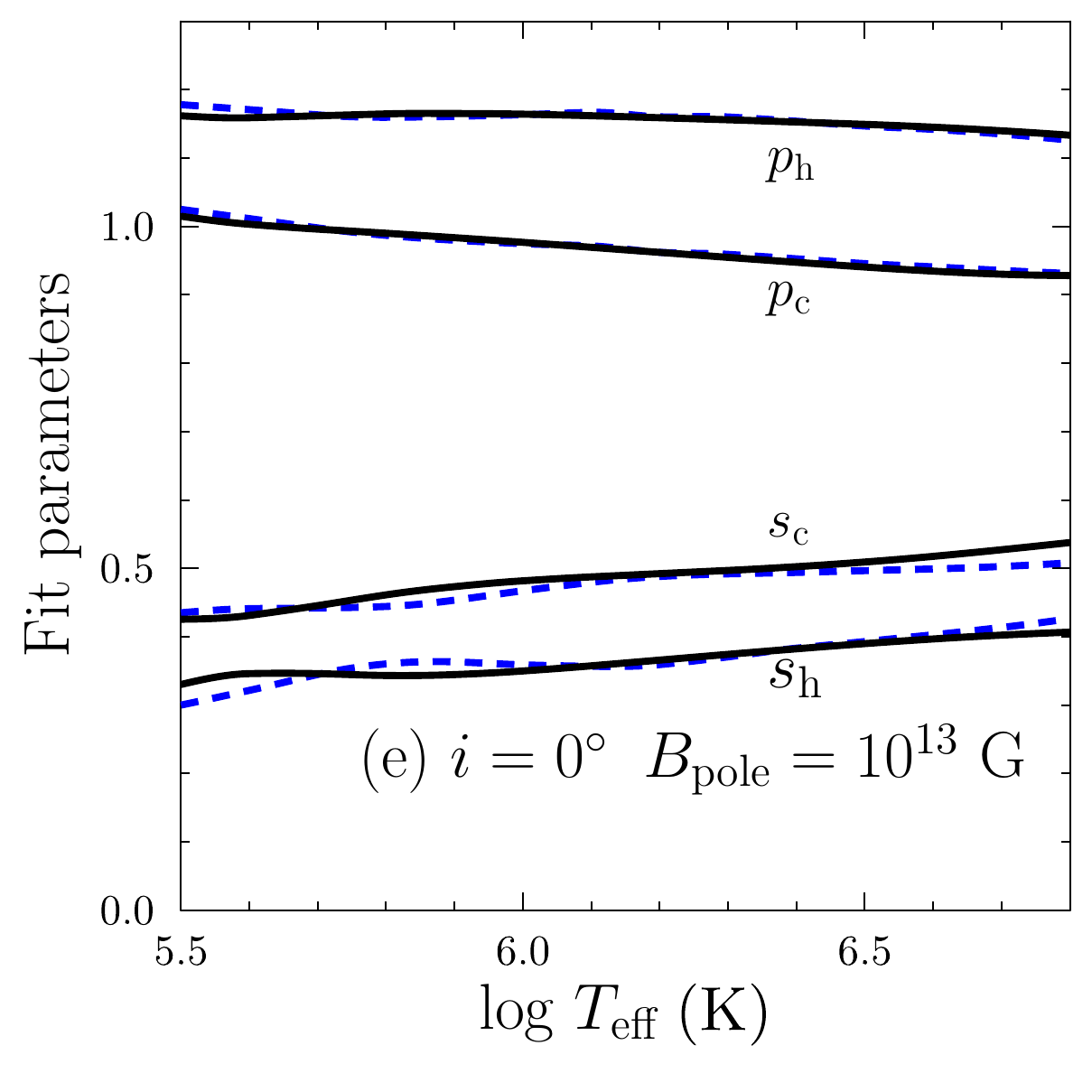}%
    \hspace{-7mm}
    \includegraphics[width=0.27\textwidth]{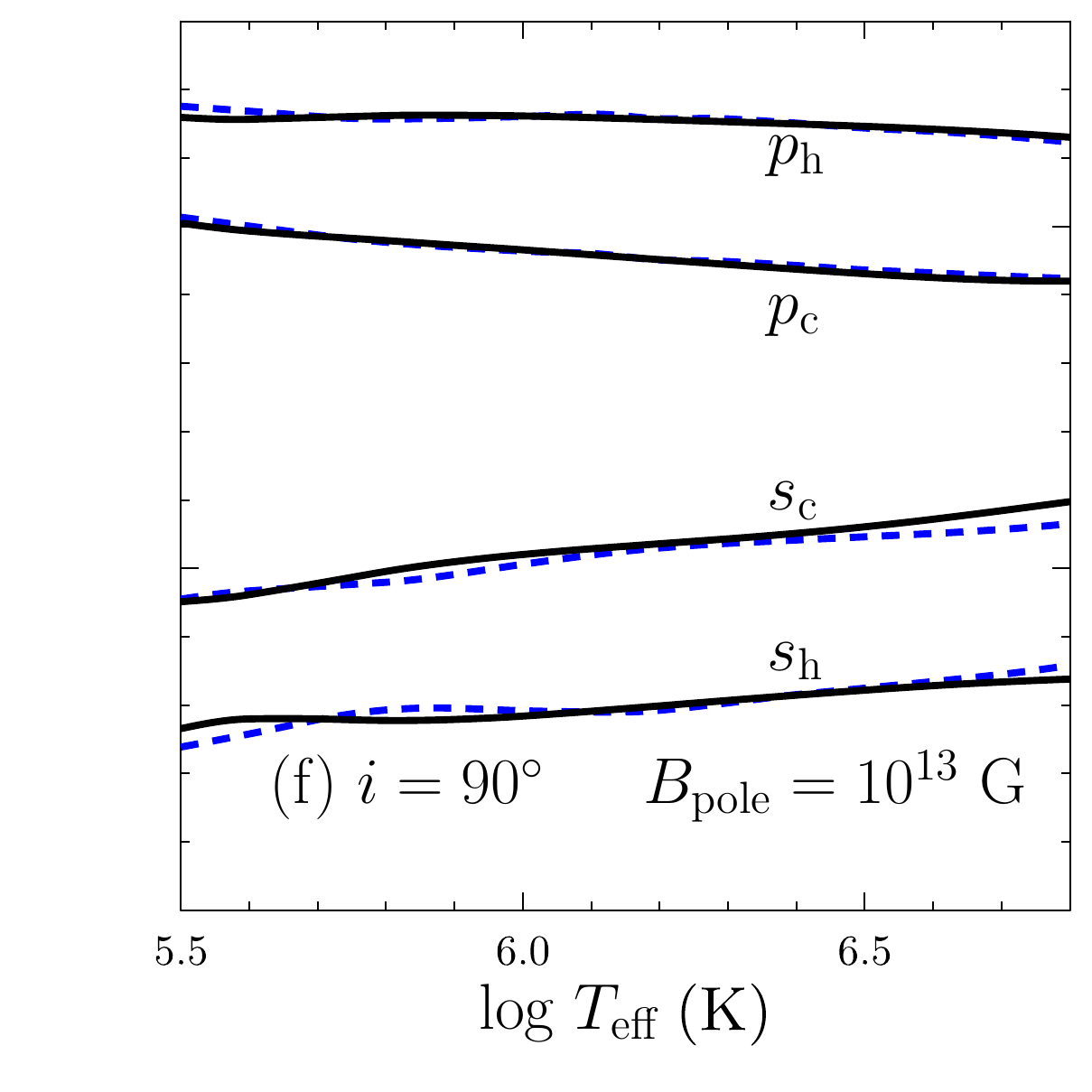}%
    \hspace{-5mm}
    \includegraphics[width=0.27\textwidth]{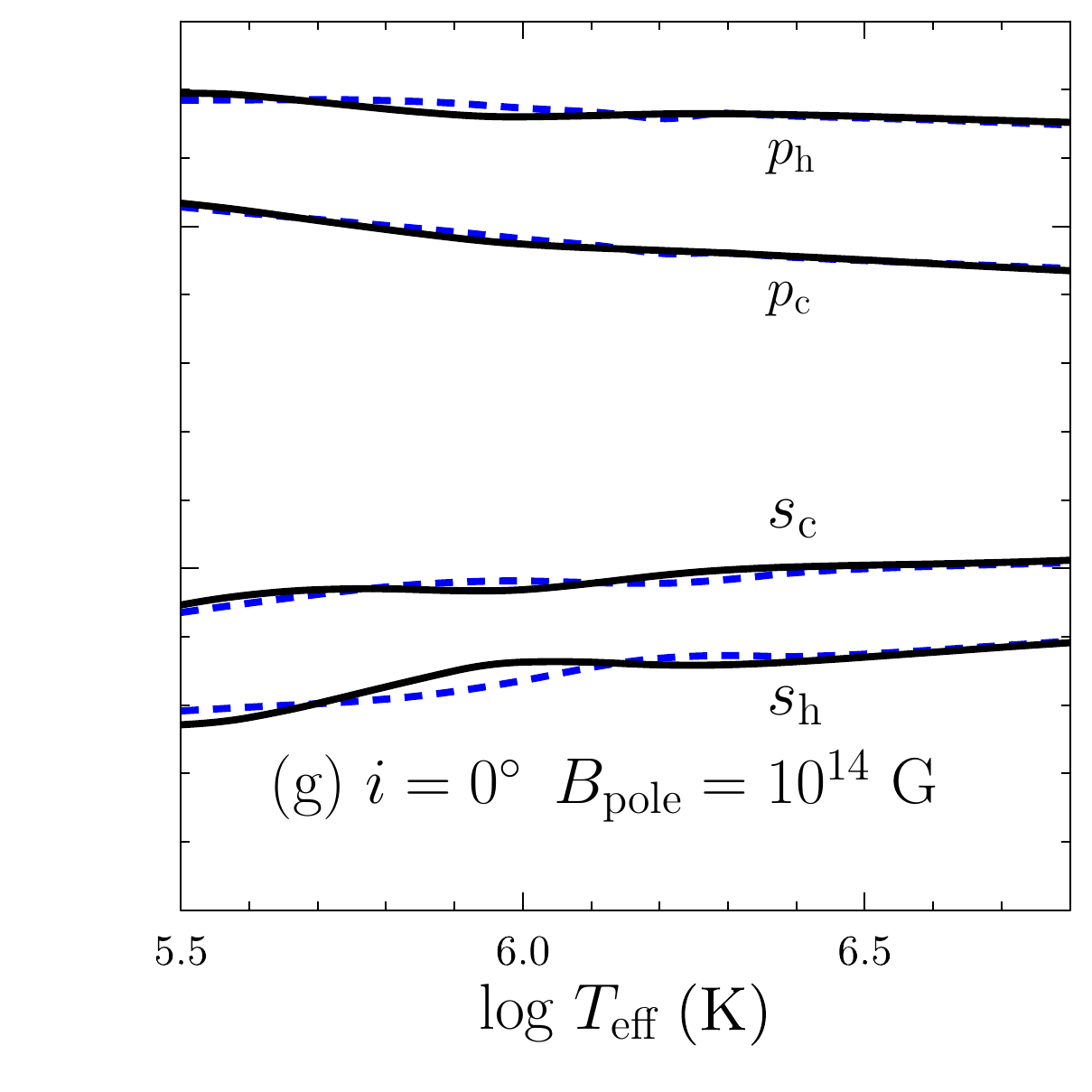}%
    \hspace{-7mm}
    \includegraphics[width=0.27\textwidth]{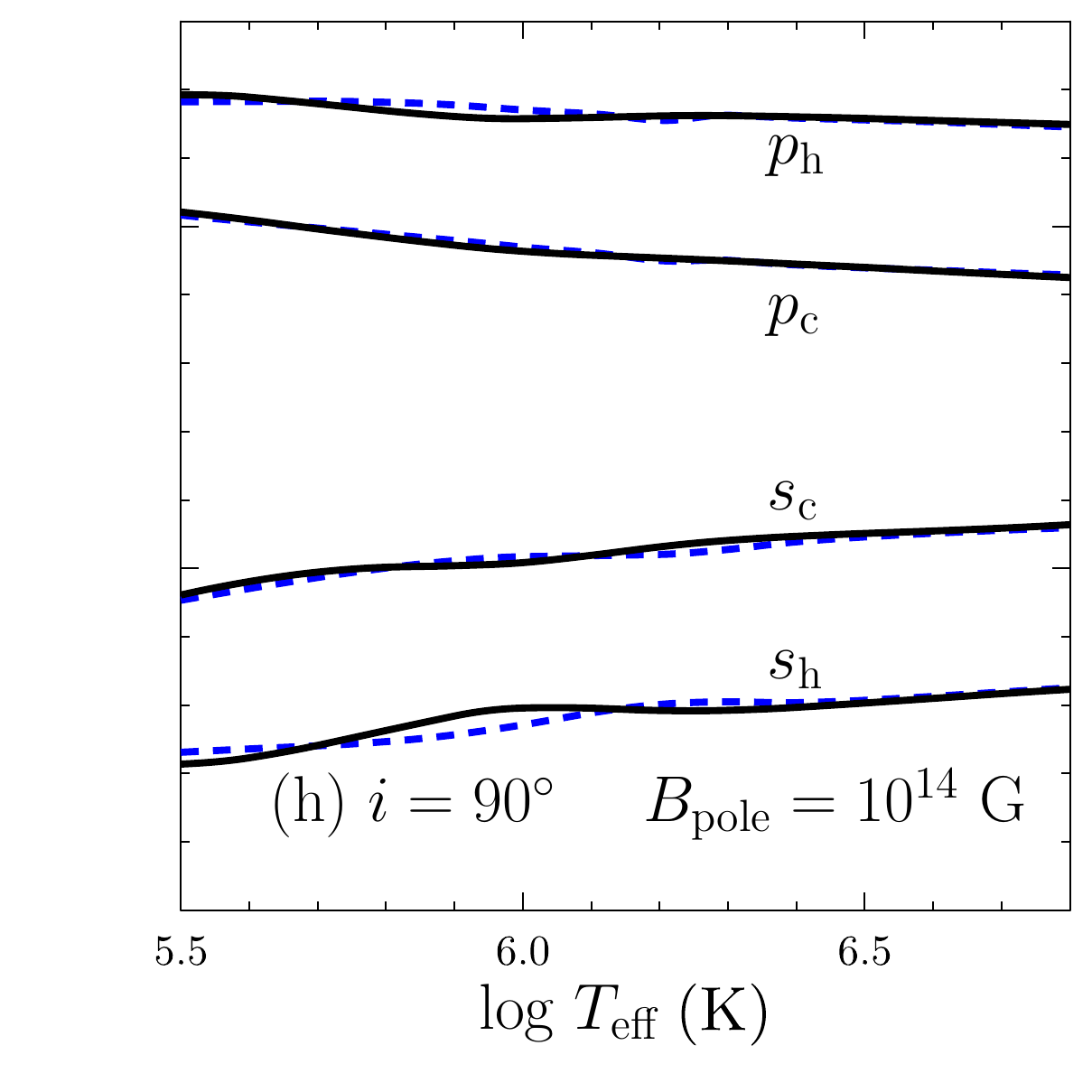}%
    \\
    \vspace{3mm}
	\caption{
		2BB fit parameters versus  
		$\log \Teff$ for
		a $1.4\,\msun$ neutron star with $R=12$ km and $\Bpole=10^{11}$, $10^{12}$, $10^{13}$, 
		or $10^{14}$~G [panels (a--h)] for polar ($i=0$)
		and equator ($i=90^\circ$) observations. 
		Solid lines refer to the star with iron heat
		blanket, while dashed ones are for accreted heat
		blanket.
	}
	\label{f:maps}
\end{figure*}

All three fluxes for each $\Teff$ look
close in the logarithmic format of Fig.\ \ref{f:Hparallel}.
However, they do not coincide (as clearly seen
from Fig.\ 3 of Paper I). Note that the difference
between the fluxes emitted from the star with accreted and
iron envelopes is noticeably smaller than the difference
between these fluxes and 1BB ones. The difference
from 1BB fluxes monotonically increases with
growing $\Bpole$ and $E$, which is quite understandable.
Higher $\Bpole$ produces stronger anisotropy of
$\Ts$ distribution. Radiation at higher energies is 
predominantly emitted from hotter places of the surface.

Let us remark that at $\Bpole=10^{13}$ G the spectral fluxes
from the star with iron and accreted heat blankets are 
almost indistinguishable (but sufficiently different
from the 1BB flux). This is because the $\Ts$ distrbutions
for iron and accreted blankets are very close
(Fig.\ \ref{f:Ts}, Sect.\ \ref{s:Palex}).
They differ only in cold equatorial belts, but the contribution
into the fluxes from
cold narrow belts appears almost neglegible
(also see Paper I). In contrast, the contrbution from
hotter surface regions can be important (Sect.\ \ref{s:hotspots}). At 
$\Bpole =10^{14}$~G the situation remains 
nearly the same as at $10^{13}$~G.

Fig.\ \ref{f:Hperp} shows similar spectral
fluxes but for equator observations (and only
for $\Bpole$= $10^{11}$ and $10^{13}$ G). 
In this case the situation is similar
to that for pole observations.

The main outcome of Figs. \ref{f:Hparallel} and 
\ref{f:Hperp} is that the difference of the
fluxes emitted from stars with iron and accreted
heat blanketing envelopes is rather small and can 
be ignored in many applications, especially 
taking into account approximate nature of heat
blanketing models. The results
for partly accreted heat blankets would be similar.
Accordingly, one can neglect chemical composition of heat
blankets for calculating thermal emission from
magnetized neutron stars in the adopted model 
of heat blankets. Therefore, the chemical composition
of heat blankets does affect cooling of the neutron stars
but almost unaffects thermal emission (under formulated
assumptions). 

\subsection{Phase-space maps}
\label{s:maps}

As discussed in Sect.\ \ref{s:2BBimage} (see also Paper I),
thermal spectral fuxes can be presented by maps of
fit parameters ($p_{\rm c},\, p_{\rm h},\, s_{\rm c},\ s_{\rm h}$) as functions of
input parameters. For completeness,  Fig.\ \ref{f:maps}
shows these 2BB maps versus log$\Teff$ at $\Bpole=
10^{11}$~G [panels (a) and (e)], $10^{12}$~G 
[panels (b) and (f)], $10^{13}$~G [(c) and (g)] and
$10^{14}$~G [(d) and (h)]. Panels (a)--(d) 
refer to pole observations, while (e)--(h) 
to equator observations. Solid lines correspond to
the iron heat blankets, while dashed lines to the accreted ones.

One can see that the dependence of fit parameters
on $\Teff$ is smooth, without any specific
features. The parameters of both effective BB components
are of the same order of magnitude. Most importantly, the
temperature $T_{\rm effh}$ of the hotter BB component
exceeds the temperature $T_{\rm effc}$ of the
colder component by less than 20 \%. The effective 
surface fraction $s_{\rm c}$ of the colder component
slightly exceeds the fraction $s_{\rm h}$ of the hotter
component.

The maps in Fig.\ \ref{f:maps} represent the same 
thermal fluxes, which are plotted in Figs.\ \ref{f:Hparallel}
and \ref{f:Hperp}. The main conclusion of Sect.\ \ref{s:fluxes}, 
based on Figs.\ \ref{f:Hparallel}
and \ref{f:Hperp}, is that the fluxes for iron and accreted
envelopes are nearly the same. If they were identical, 
then respective solid and dashed curves in 
Fig.\ \ref{f:maps} should have 
been identical as well. However, these curves differ.
The difference is especially visible for $\Bpole=10^{11}$
and $10^{12}$~G, but less visible at higher $\Bpole$. The
difference is mainly explained by two things. First,
weak variations of 2BB fit parameters lead to
weaker variations of the fluxes. Second, the 2BB approximation
is accurate but not exact by itself, as discussed in the
previous section.

It is important that we fit {\it unabsorbed}
calculated spectral fluxes using {\it unabsorbed} 2BB models. 
Accordingly, as argued in Paper I, one can compare 2BB
parameters, inferred from observations (corrected for
interstellar absorption and non-thermal radiation component),
with the theoretical parameters of unabsorbed 2BB models.    
   
%
%

\section{Extra heating of magnetic poles}
\label{s:hotspots}

\begin{figure*}
	\centering
	\includegraphics[width=0.4\textwidth]{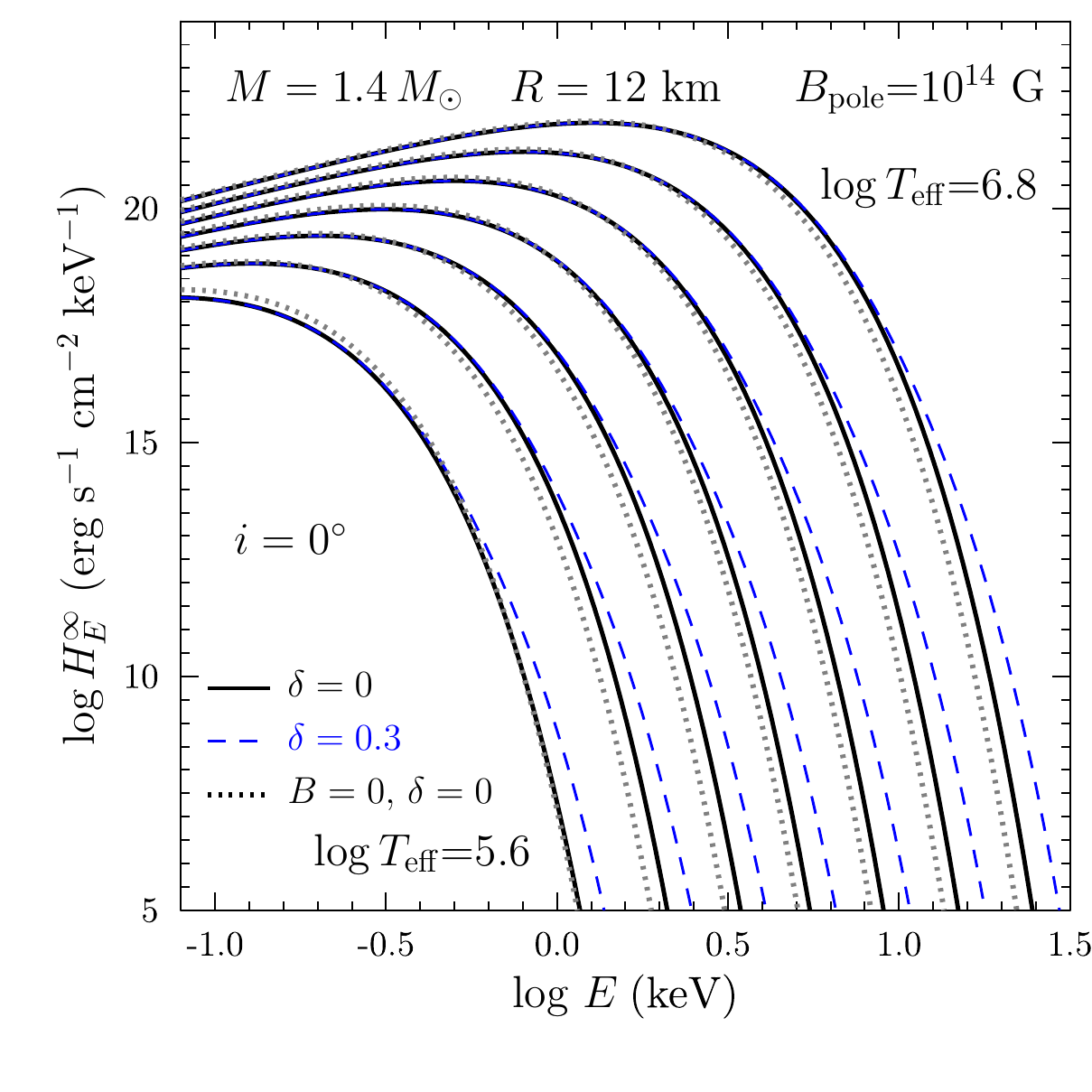}%
	\includegraphics[width=0.4\textwidth]{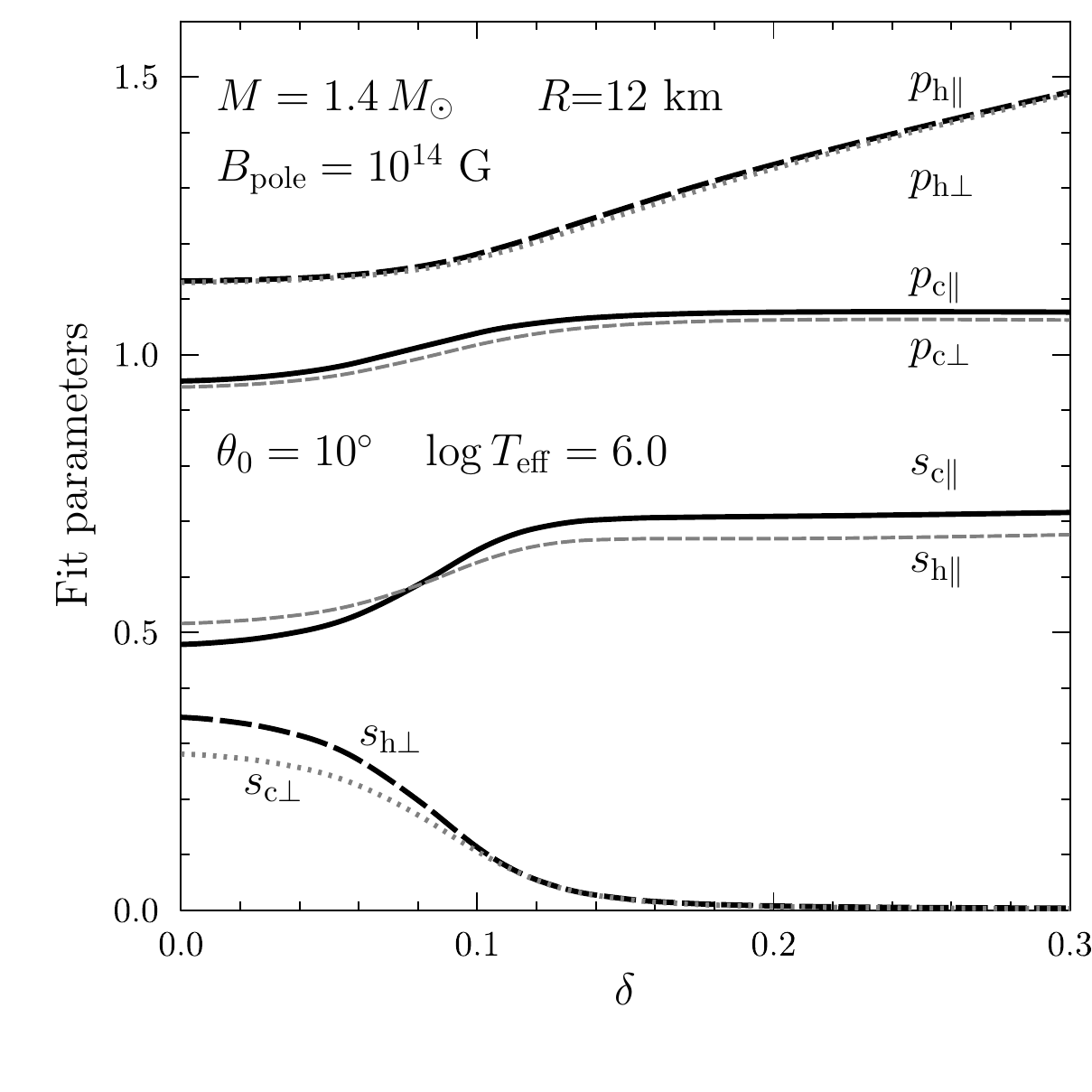}%
	\caption{
		{\it Left:} Thermal spectral fluxes observed from magnetic poles 
		of a neutron star with 
		$M=1.4\,\msun$ and $R=12$ km at 
		$\Bpole=10^{14}$~G  and $\log \Teff~$[K]=
		5.6, 5.8,\ldots,6.6. 
		The solid lines are obtained without extra
		heating, while the dashed lines 
		are for two extra hotspots of angular size
		$\vartheta_0=10^\circ$ on magnetic poles 
		at the heating 
		parameter $\delta=0.3$. Dots
		show 1BB model for a non-magnetic star to guide the eye. 
		 {\it Right:}	
		2BB fit parameters for the same star with
		$\Teff$=1 MK  
		possessing two polar hotspots 
		of angular size
		$\vartheta_0=10^\circ$ on magnetic poles [see Eq.\
		(\ref{e:Ts})] versus extra relative surface
		temperature increase $\delta$ at the pole.
	}
	\label{f:spect2hotspots}
\end{figure*}

Sect.\ \ref{s:acc} describes thermal emission of cooling
neutron stars with dipole magnetic field near their surfaces,
using the $\Ts$ model 
calculated in \cite{PY01,PYCG03}. According to
the theory, radiative spectral
fluxes should be almost 
insensitive to the chemical
composition of the heat blanketing envelopes; 
2BB fits to their thermal X-ray spectral fluxes do not 
give the difference between the temperatures
$T_{\rm effh}$ and $T_{\rm effc}$ of the hotter
and colder BB components higher than 20\% (for $\Bpole
\lesssim 10^{14}$~G). 

A brief comparison with observations of isolated middle-aged neutron stars in Paper I shows that there are no
reliable candidates for such objects at the moment; the difference 
between $T_{\rm effh}$ and $T_{\rm effc}$ is actually 
larger, at least $\sim 50\%$ for RX J1856.5$-$3754
\cite{Sartore_ea12,Yoneyama_ea17}, as
a promising example.

Here we propose a possible phenomenological
extension of the discussed model, which
may simplify interpretation of observations
of some sources. Specifically, let us assume
the presence of additional heating of magnetic
poles, which produces
polar hotspots and raises the pole temperature. Such
an extension has been used, for instance, 
in Paper I but assuming heating of both poles.
Now we consider heating either of both poles
or one of them. For simplicity,  
this heating is assumed to be axially symmetric.
It does not violate the axial symmetry of the
$\Ts$ distribution. Note that two equivalent
hotspots do not destrpy the initial symmetry of
the north and south hemispheres, whereas
one hotspot does destroy it. Possible physical 
justifications
for extra heating are mentioned in Sect. \ref{s:discuss}.

Let $T_\mathrm{s0}(\vartheta)$ be the basic effective 
surface temperature used in
Sect.\ \ref{s:model}. We introduce
a small phenomenological angle $\vartheta_0$ 
that deteminies the size of 
a hotspot. Following Paper I we assume that
at $\vartheta < \vartheta_0$ (in the nothern hotspot),
\begin{equation}
   \Ts(\vartheta)=T_\mathrm{s0}(\vartheta)\,
   \left[1+\delta \, \cos^2 \left( \frac{\pi \vartheta}{2\vartheta_0} \right)\right]
   \quad \mathrm{at}~\vartheta \leq \vartheta_0,
   \label{e:Ts}
\end{equation} 
and $\Ts(\vartheta)=T_{\rm s0}(\vartheta)$ outside
the hotspot.  The parameter $\delta$
specifies an extra temperature enhancement at the magnetic pole; the enhancement 
smoothly disappears as $\vartheta \to \vartheta_0$. 
This will be our {\it model with one hotspot}.
Assuming similar temperature enhancement near the second 
magnetic pole, we will get the {\it  model with
two hot spots}.
The presence of spots 
renormalizes the total effective temperature $\Teff$, Eq.~(\ref{e:Ls}). 

Otherwise computation of spectral fluxes is
the same as in Sect.\ \ref{s:model}.
These fluxes can also be approximated by 2BB fits 
(\ref{e:2BBfit}), and 
can be analyzed via phase-space maps. In some cases
presented below 
the relative the fit accuracy becomes worse (reaching sometimes $\sim 10\,\%$) but the fit remains 
robust. We will again take the star with
$M=1.4 \, \msun$ and $R=12$ km. For certainty, 
we use the model of iron heat blanket and
$\vartheta_0=10^\circ$. 

\subsection{Extra hotspots on both poles}
\label{s:2hotspots}

The left panel of Fig.\ \ref{f:spect2hotspots} 
demonstrates calculated spectral
fluxes for $\Bpole=10^{14}$~G and seven values of
$\log \Teff$[K]=5.6, 5.8,\ldots, 6.8 at $\delta=0.3$.
Since these hotspots are identical, 
observations of the first and second poles ($i=0$ and $180^\circ$) give the same fluxes. 
Such fluxes are shown by dashed lines. For comparison, the solid lines are calculated at $\delta=0$, in which case 
the extra hotspots are absent and the results of Sect.\ 
\ref{s:model} apply. The dotted lines show the fluxes
for non-magnetic star. One can see that the presence of hotspots 
enhances the spectral fluxes at high photon energies.
This is expected: enhanced surface temperature of hotspots
intensifies generation of high-energy radiation.
The higher $\Teff$, the larger photon energies are
affected. 
The solid and dotted lines are the same as in Fig. 3c.

The right panel of Fig.\ \ref{f:spect2hotspots} 
shows the 2BB fit parameters versus $\delta$ for the same star 
with $\Teff$=1 MK. 
At $\delta=0$,
the results naturally coincide with
those in Figs.\ \ref{f:maps}g and \ref{f:maps}h. 
However with increasing
$\delta$, the fit parameters become different. The temperature $T_{\rm effh}$ of the hotter
BB component  and the effective 
emission surface area $s_{\rm c}$ of the colder component noticeably increase,
whereas the emission surface area of the hotter component dramatically
falls down. Even with really small hotspot temperature enhancement
$\delta \leq 0.3$ (which gives the 
fraction of extra luminosity $\leq$1.4\%), one gets absolutely new 
phase-space portrait with $s_{\rm h} \ll s_{\rm c}$. Such 2BB fits
have been often inferred from observations of cooling isolated neutron stars
(see, e.g., Paper I); these sources are usually interpreted
as neutron stars, which have small hotspots with noticeably enhanced temperature. 

Therefore, the theory with hotspots predicts 
(e.g. Paper I) two types of 
neutron stars, whose spectra can be approximated 
by the 2BB models.
The first ones are those with smooth surface temperature
distributions, created by nonuniform surface magnetic fields 
(as considered in Sect.\ \ref{s:model}) to be called
spectral 2BB models  {\it of smooth magnetic atmospheres}.
The second sources are those with distinct hotspot
BB component to be called 2BB
{\it with hotspots}. Naturally, there is 
a smooth transition between them
(for instance, by increasing $\delta$ in the right
panel of Fig.\ \ref{f:spect2hotspots}). It seems that the 
observations do not provide good
candidates for the sources of the first type (Paper I),
but there are some candidates for the sources of the second and
intermediate types  as we outline later.

\subsection{Phase resolved spectroscopy in case of two hotspots}
\label{s:phaseresove2hotspots}

\begin{figure*}
	\centering
	\includegraphics[width=0.4\textwidth]{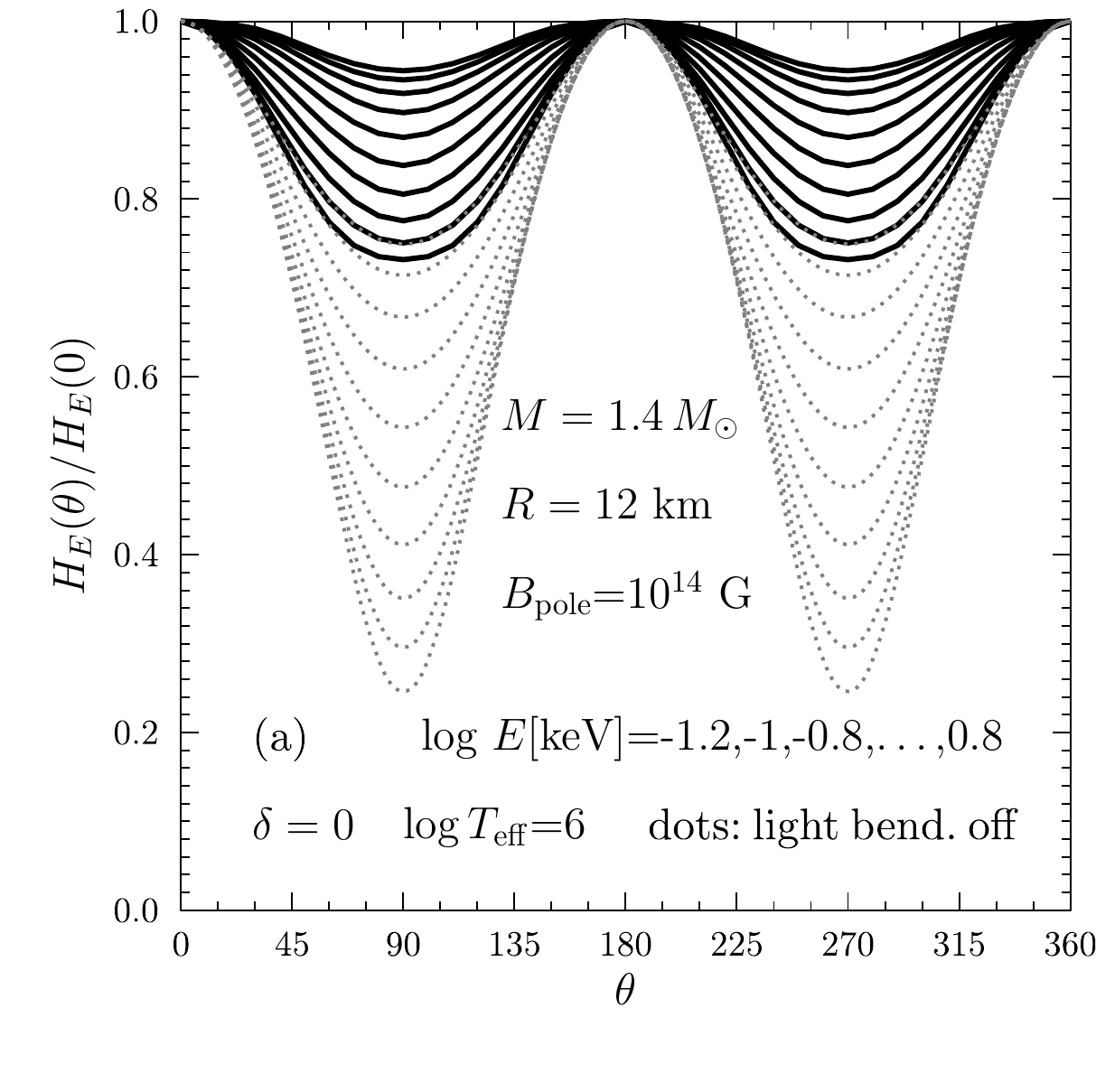}%
	\hspace{-8mm}
	\includegraphics[width=0.4\textwidth]{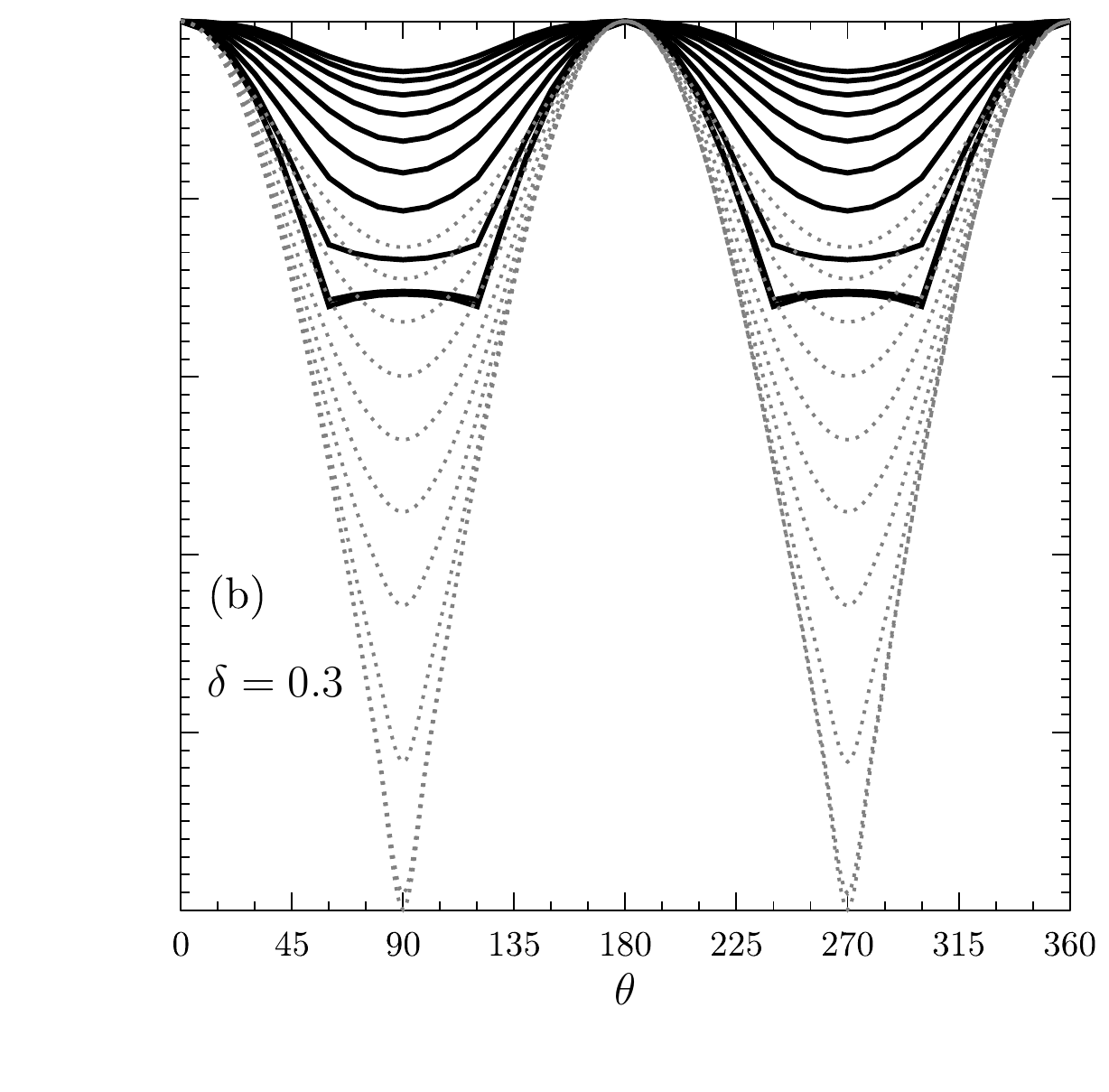}%
	\caption{
		Theoretical lightcurves  normalized to pole
		observations versus phase angle $\theta$ 
		for a neutron star with
	    $\Teff$=1~MK and $\Bpole= 10^{14}$ G; (a) $\delta=0$ (no extra pole heating) and 
	    (b) $\delta=0.3$. 
	    Solid  curves are for photon energies $\log E$[keV]=
		$-1,\,-0.8,\ldots,0.8$ (from top to bottom). 
		The two lowest curves on panel (b) (with highest $E$)
		almost coincide. 
		Dotted curves show the same but neglecting gravitational light
		bending. The star is assumed to be
		orthogonal rotator with the spin axis perpernicular
		to the line of sight. 	
	}
	\label{f:phase2hotspots}
\end{figure*}

Let us present a few calculated lightcurves to illustrate an important problem of pulse fraction. For simplicity, we consider
the star as orthogonal rotator with the spin axis
perpendicular to the line of sight. Fig.\ \ref{f:phase2hotspots}
presents several lightcurves for a star with
$\Bpole=10^{14}$~G versus phase angle $\theta$. The lightcurves 
are normalized by $H_E(0)$ (the spectral flux for pole 
observations at given energy $E$). The displayed ratio of
redshifted or non-redshifted fluxes is the same at a given
$E$, so that we drop the symbol $\infty$. Note  that the normalization flux
$H_E(0)$ strongly depends on energy by itself. 
Panel (a) corresponds to $\delta=0$, in which case
the hotspots are actually absent and we have the emission
from the smooth magnetic atmosphere. In case (b) 
the hotspots with $\delta=0.3$ are available and
strongly affect the lightcurves.

Solid lines on each panel in Fig.\ \ref{f:phase2hotspots} 
present the lightcurves at 10 energies (from top to
bottom), $\log E$[keV]$=-1, -0.8, \ldots, 0.8$. The higher
$E$, the stroger phase variations.  The last two
lines on panel (b) 
almost coincide. 
The lightcurves on panel (a) are
seen to be smooth. The curves on panel (b) at $E \lesssim 1$ keV
are although smooth and resemble those on panel (a).
However, at higher energies the lightcurves (b) have
shapes with erased dips at nearly equator observations 
[and then Eq.\ (\ref{e:Hanyi}) becomes inaccurate although 
Eq.\ (\ref{e:2BBfit}) works well]. These
curves are typical for lightcurves produced
by antipodal point sources on the neutron star surface;
see, e.g. Fig.\ 4 in \cite{2002Beloborodov} or Fig.\ 6 in \cite{2020Poutanen}. Clearly, at high energies 
the star emits from the poles,
the hottest places on the surface. 
Nevertheless, the pulse fraction remains
not too high ($\leq 30\%$) in both cases (a) and (b)
even at the highest taken energy $E \approx 6.3$ keV,
at which thermal spectral flux becomes typically very low
by itself.
Therefore, the enhancement of thermal
emission by extra heating of two magnetic poles does
not lead to sizable pulse fractions. 

The dotted lines on Fig.\ \ref{f:phase2hotspots} show
the same lightcurves as the solid lines but calculated
neglecting gravitational bending of light rays. 
One sees that without the light bending the
pulse fraction would be much stronger 
(as is well known; see e.g. \cite{page95}).
The difference of solid and dotted curves on Fig.\ \ref{f:phase2hotspots} has simple explanation. Without
light bending, the equator observations ($\theta=90^\circ$ 
or 270$^\circ$) would collect more light emitted from the
cold equator. This would lower the observed emission 
and produce stronger dips of the lightcurves. 

\subsection{Extra hotspot on one pole}
\label{s:1hotspot}

\begin{figure*}
	\centering
	\includegraphics[width=0.4\textwidth]{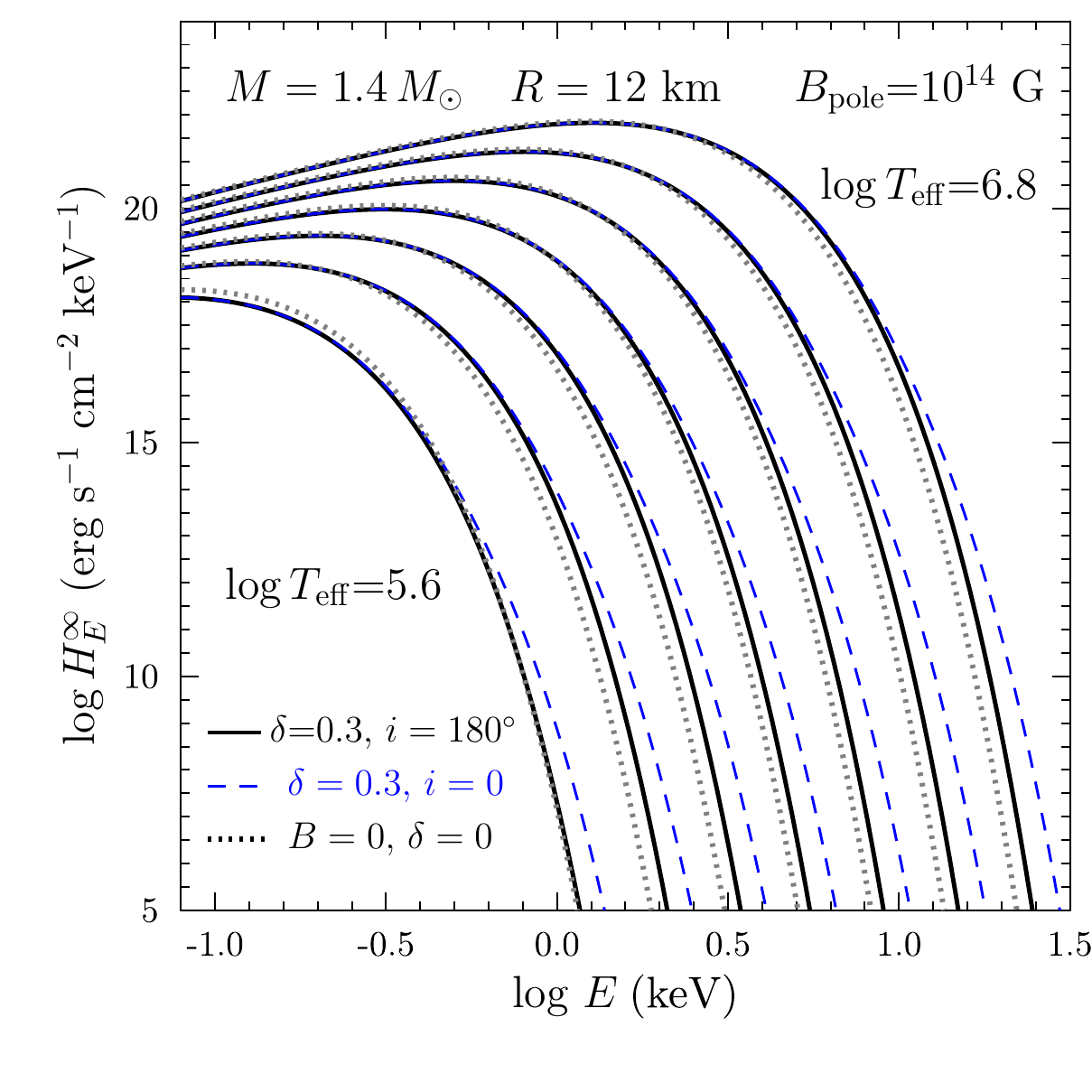}%
	\includegraphics[width=0.4\textwidth]{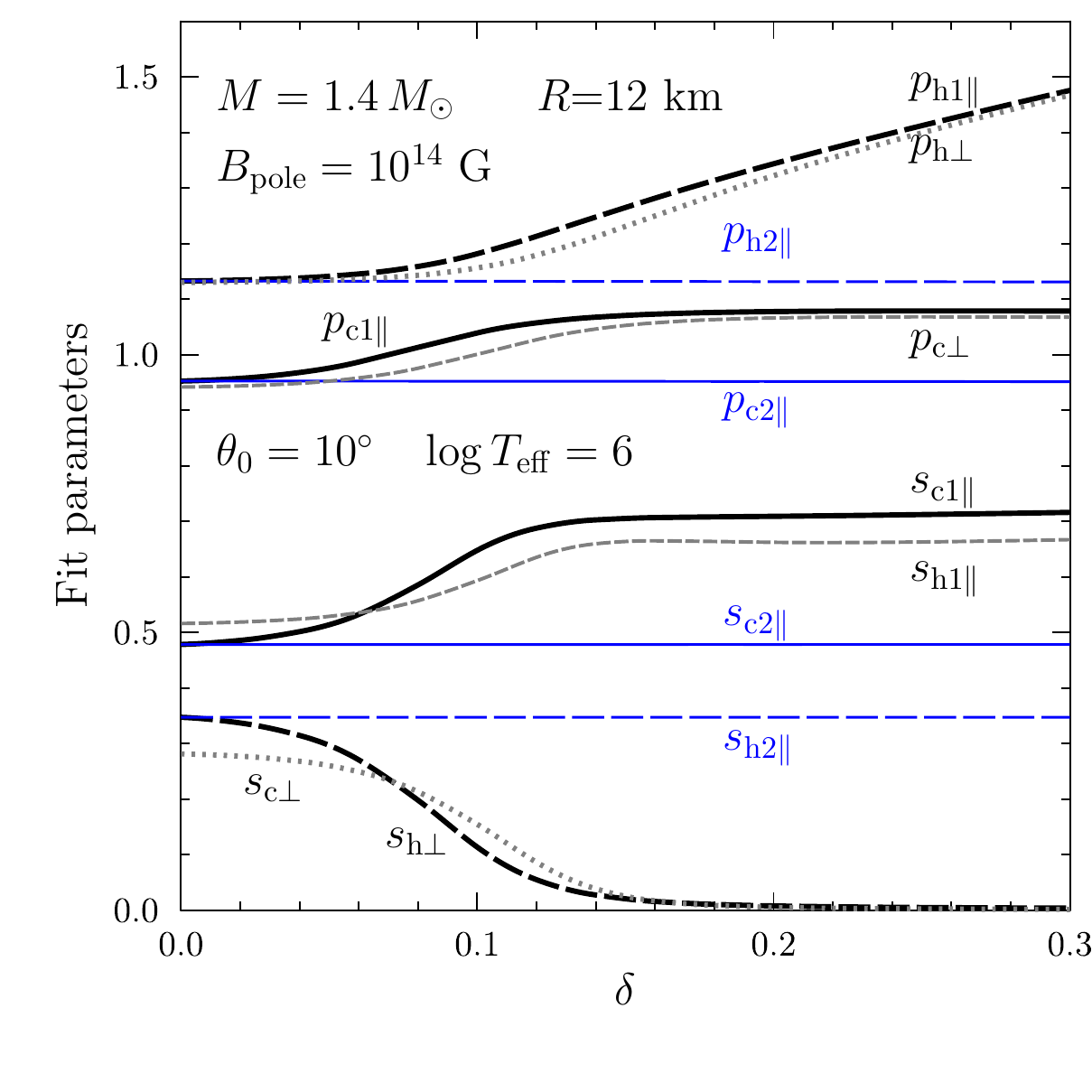}%
	\caption{
		{\it Left:} Theoretical spectral fluxes for a neutron star with
		$M=1.4\,\msun$, $R=12$ km, $\Bpole= 10^{14}$ G,
		different $\Teff$ and one extra hotspot on one magnetic pole 
		at the extra heating parameter 
		$\delta=0.3$. 
		The solid and dashed curves are for
		non-heated ($i=180^\circ$) and heated pole ($i=0$) observations, respectively.
		{\it Right:}
		Maps of fit parameters versus $\delta$ for this star at
        $\Teff$=1 MK.
       The thick solid and gray curves are for the
       north-pole and equator observations, respectively. Thin
       horizontal curves are for the south-pole
       observations. 		 		
	}
	\label{f:spect1hotspot}
\end{figure*}

Finally, consider the case of a single hotspot,
which we put at the north pole. This
case is different from the previous one. 

The left panel of Fig. \ref{f:spect1hotspot} presents spectral
fluxes calculated for the same values of $\Teff$ and
$\Beff$ as in Fig. \ref{f:spect2hotspots}. The dashed
lines are for the north-pole observations 
($i=0$) at $\delta=0.3$. These fluxes are
the same as in Fig. \ref{f:spect2hotspots}: the observer
sees a large fraction of the surface
(mainly the northern hemisphere);
the southern polar region is unseen. 

The solid line in the left panel of 
Fig.\ \ref{f:spect1hotspot} refers
to the same case of $\delta=0.3$ but the star is
observed from the south pole ($i=180^\circ$). 
Then the observer cannot see the hotspot, 
so the line exactly coincides with the solid line in Fig.\ \ref{f:Hparallel}c.
The dotted line in Fig.\ \ref{f:spect1hotspot} is
for a non-magnetic star (as in Figs.\ \ref{f:Hparallel}c
and \ref{f:spect2hotspots}). Now, with only one extra heated pole, 
the difference between observations of the 
north and south poles is  pronounced.

The right panel of 
Fig.\ \ref{f:spect1hotspot} presents the maps
of 2BB fit parameters versus $\delta$ 
for the same star at $\Teff=1$ MK  
for the north-pole, south-pole, and equator observations
(to be compared with Fig.\ \ref{f:spect2hotspots}). 
The maps the for north-pole and equator observations 
are nearly the same as those in Fig.\ \ref{f:spect2hotspots},
but the maps for the south-pole observations are plain.
Corresponding fit parameters are independent of $\delta$ just 
because the observer cannot see the emission from
the hotspot (which is the only emission 
which increases with $\delta$).

\subsection{Phase resolved spectroscopy in case of one hotspot}
\label{s:phaseresove1hotspot}

\begin{figure*}
	\centering
	\includegraphics[width=0.4\textwidth]{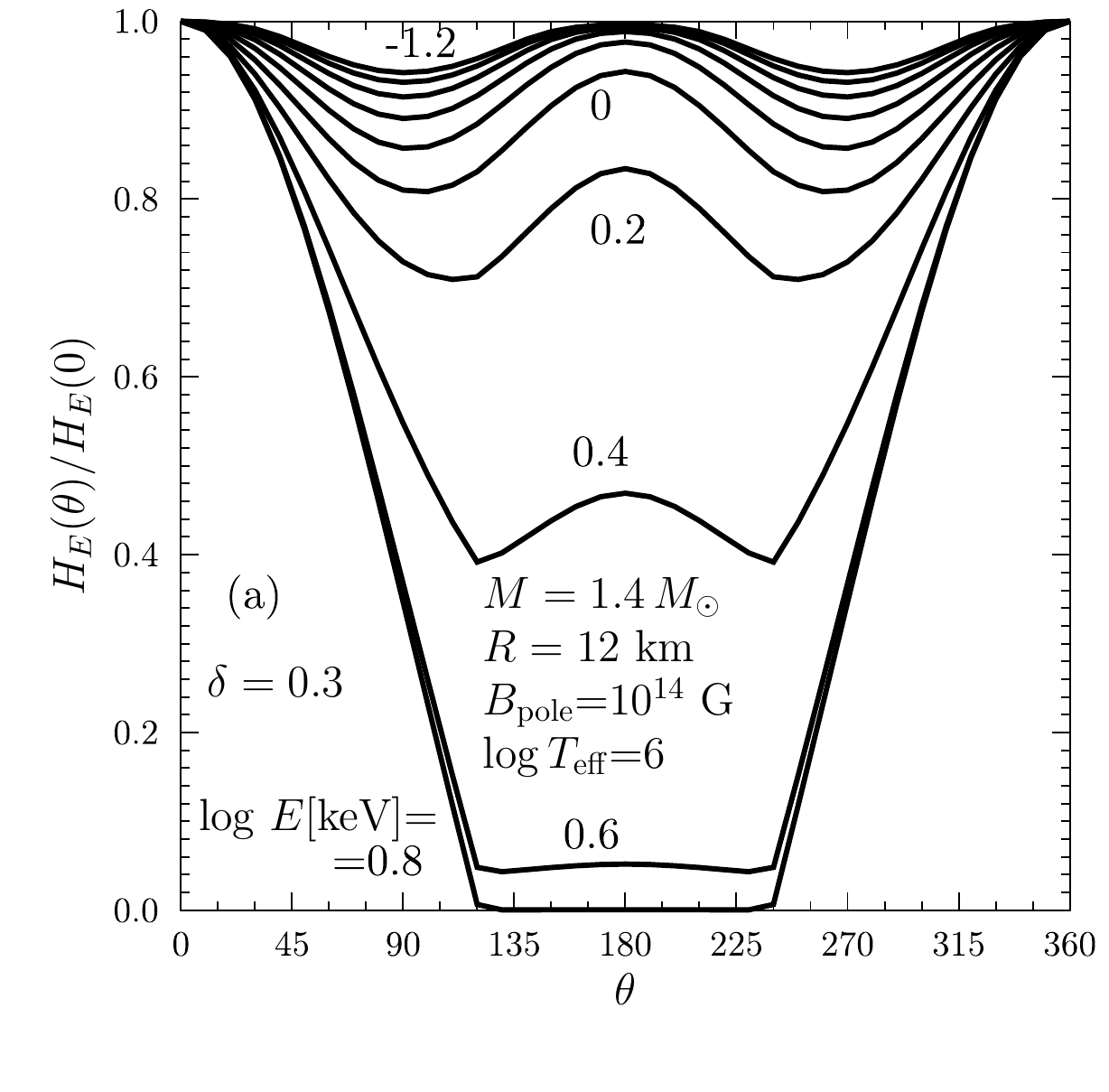}%
	 \hspace{-9mm}
		\includegraphics[width=0.4\textwidth]{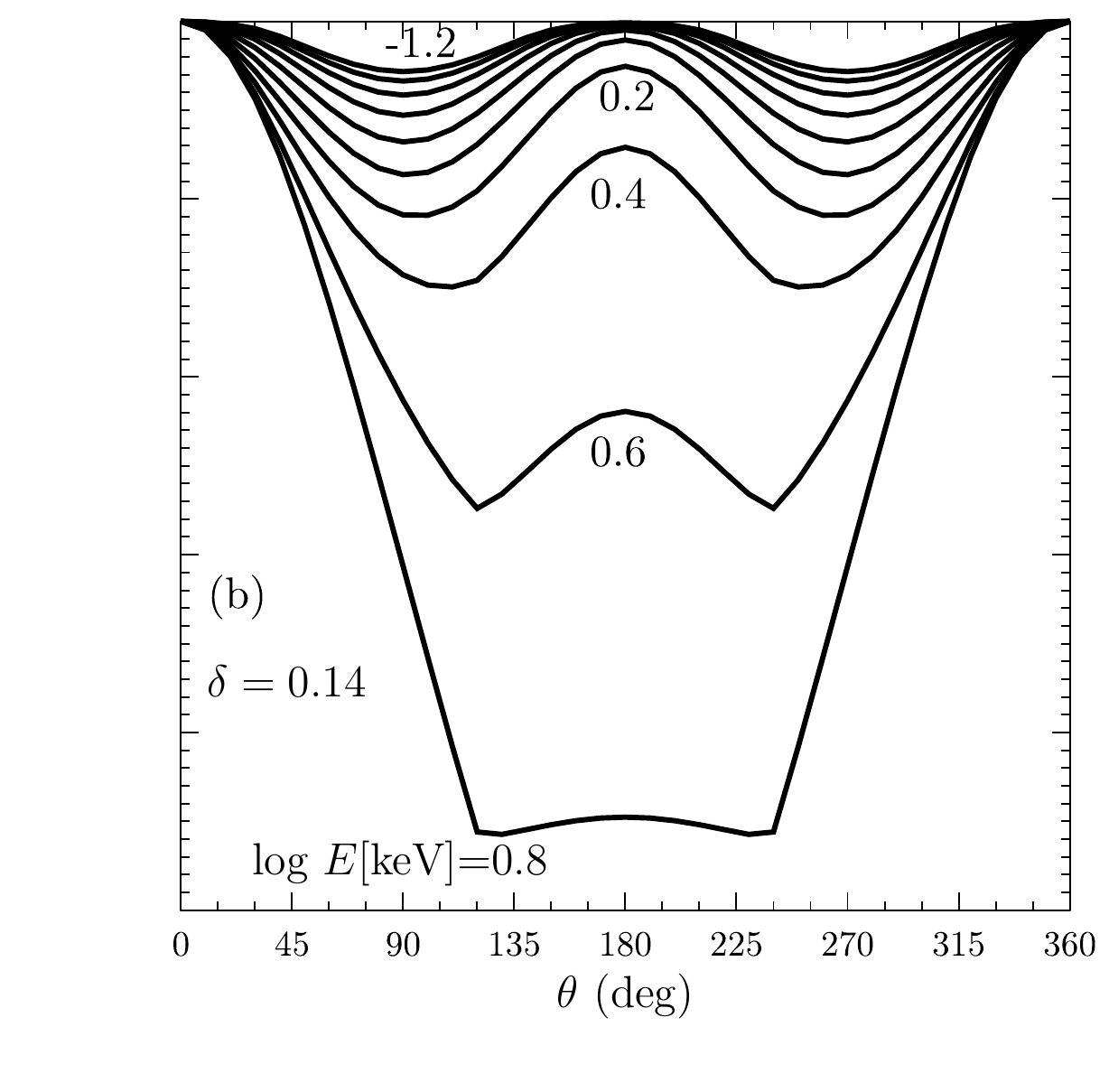}%
	\caption{
		Theoretical lightcurves for a neutron star with
		$M=1.4\,\msun$, $R=12$ km, $\Teff$=1 MK and 
		$\Bpole= 10^{14}$ G with an extra hotspot of angular size
		$\vartheta_0=10^\circ$ on one magnetic pole 
		versus phase angle $\theta$ at
		extra heating parameter 
		$\delta=0.3$ (a) and 0.14 (b). The curves are for photon energies $\log E$~[keV]=
		$-1,\,-0.8,\ldots,0.8$. 
		See the text for details.	
		}
	\label{f:pf1spot}
\end{figure*}

\begin{figure*}[t]
    \centering
	\includegraphics[width=0.4\textwidth]{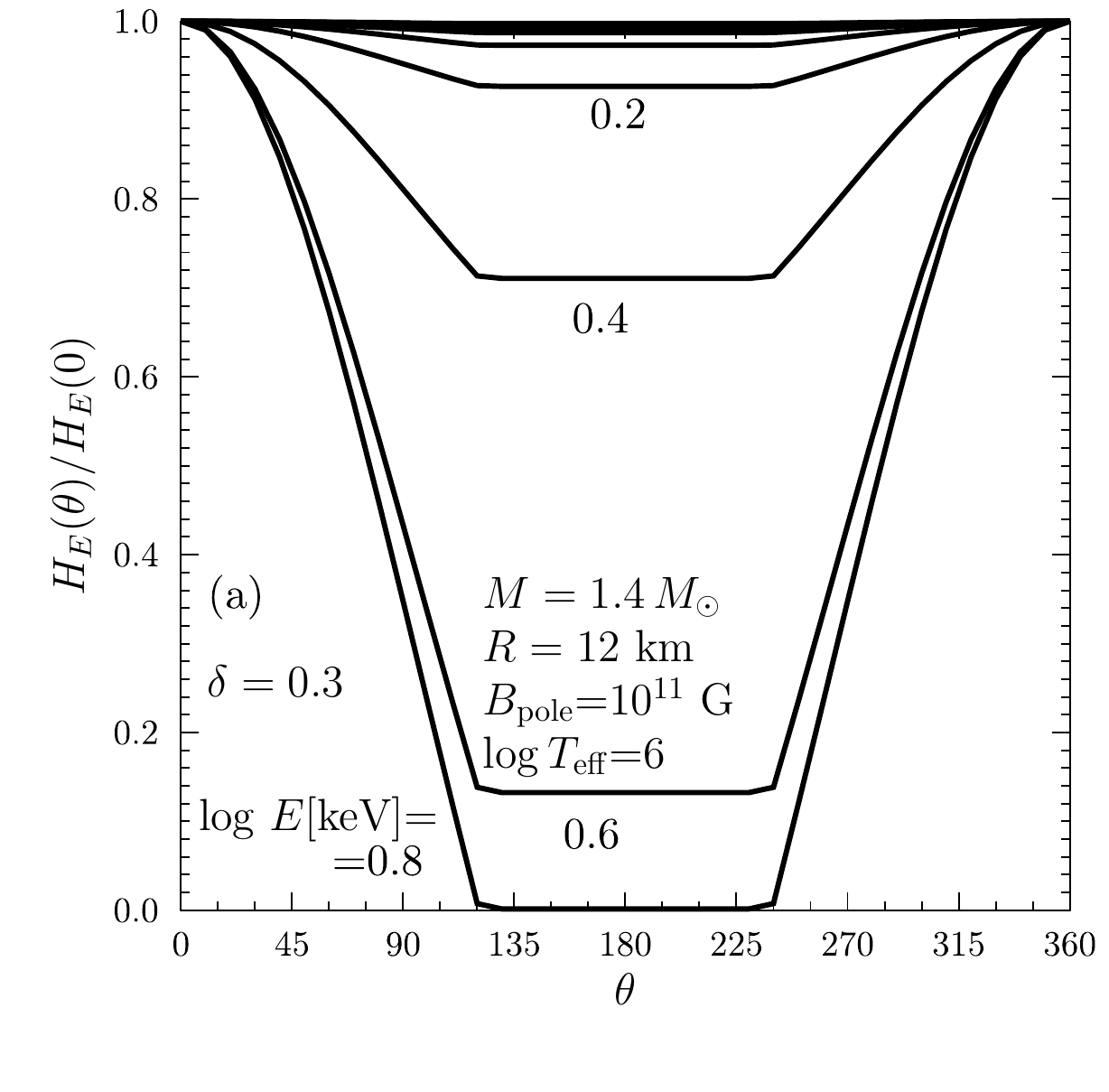}%
	\hspace{-8mm}	
	\includegraphics[width=0.4\textwidth]{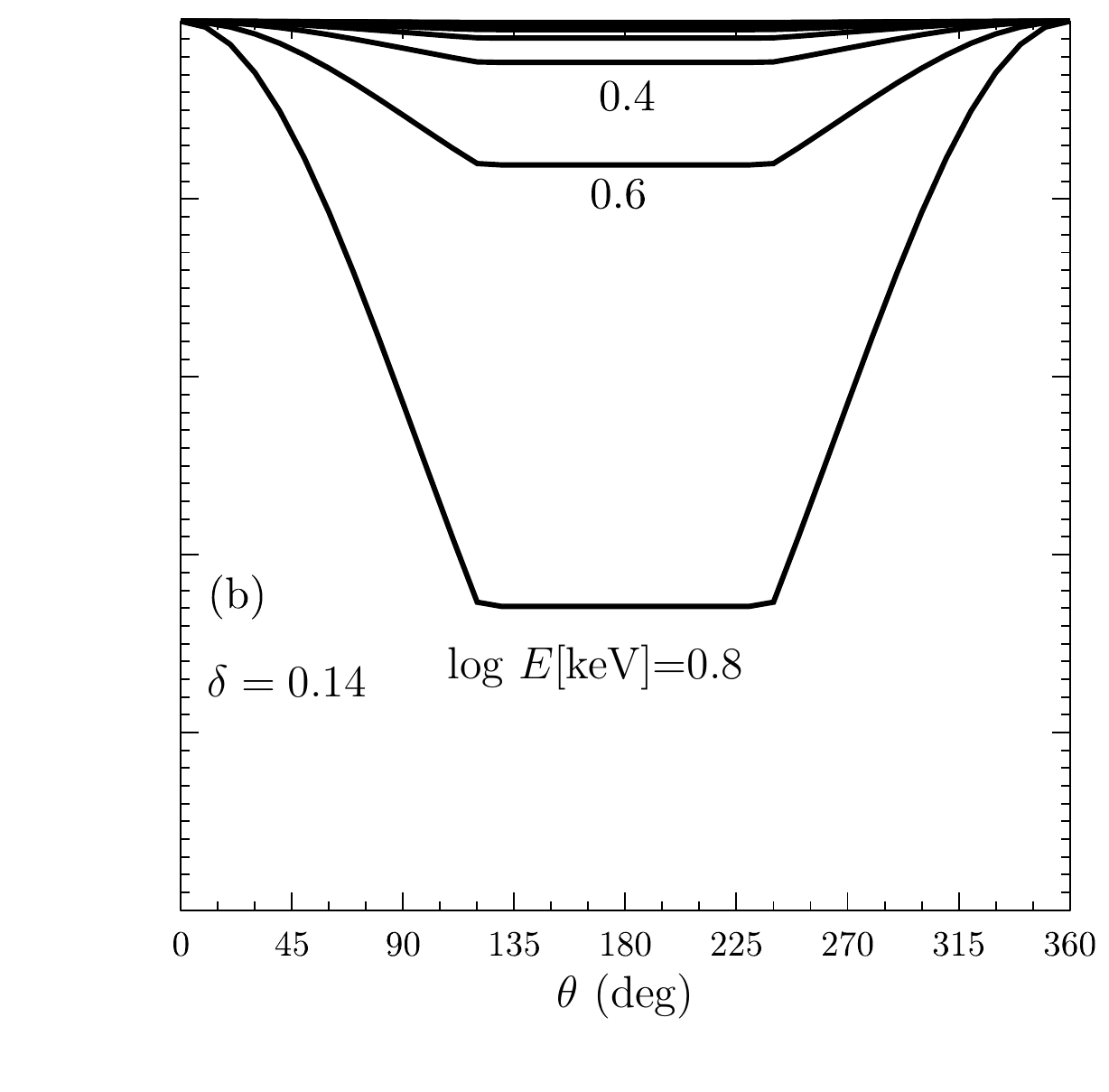}%
	\caption{
	  Same as in Fig.\ \ref{f:pf1spot} but for
	  $\Bpole=10^{11}$~G.	
	}
	\label{f:pf1spotB11del03}
\end{figure*}

Phase-resolved lightcurves are plotted 
in Figs.\ \ref{f:pf1spot} and \ref{f:pf1spotB11del03}.
They are naturally different
from the case of two extra heated magnetic poles.
Fig.\ \ref{f:pf1spot} is analogous to Fig.\ 
\ref{f:phase2hotspots} and shows the light curves
at the same $\Teff$, $\Bpole$, and extra 
heating parameter $\delta=0.3$ (a) and
$\delta=0.14$ (b). With one hotspot, the
fractions of extra luminosities are 0.7\% (a)
and 0.3\% (b).
The
phase-space variations are similar 
to those in Fig.\ \ref{f:phase2hotspots} only
at low photon energies, in which case
the extra polar heating is almost unnoticeable
and the lightcurves have two shallow 
dips over one rotation
period. However with growing $E$, the extra
heating becomes important and increases the
depth and width of the dips (such shapes
are known to be produced by point sources
on neutron star surface, e.g. Fig. 4 in 
\cite{2002Beloborodov}). At $E \gtrsim 4$ keV
in Fig.\ \ref{f:pf1spot}a
the pulse fraction becomes close to 100\%. In that case
the observer would see only one, but very pronounced 
dip over one rotation, because the extra hotspot would
be poorly vizibe in a wide range of $\theta$ 
with the center at $\theta=180^\circ$.  
In Fig.\ \ref{f:pf1spot}b  the heating is weaker,
the dips are smaller,
but nevertheless they are much larger than for
two heated spots at $\delta=0.3$ 
in Fig.\ \ref{f:phase2hotspots}. Evidently, 
one extra-heated magnetic pole is
much more profitable than two, for producing large pulse
fractions. 

Fig.\ \ref{f:pf1spotB11del03} 
is plotted for $\Bpole=10^{11}$~G to
illustrate the effects of $\Bpole$ on the pulse fraction.
Fig.\ \ref{f:pf1spotB11del03}a presents the normalized
lightcurves for the star with $\delta=0.3$. Now the pulse fraction
is lower than at $\Bpole=10^{14}$~G. Nevertheless,
it remains much higher than it would be if the additional heating
were switched on at both poles. Fig.\ \ref{f:pf1spotB11del03}b 
demonstrates that at lower $\delta=0.14$ at $\Bpole=10^{11}$~G
the pulse fraction is weaker affected by the extra heating
but nevertheless reaches about 70\% at $E \sim 6$ keV.

\section{Discussion and conclusions}
\label{s:discuss}

Several isolated middle-aged neutron stars which 
radiation spectra have been (or can be) fitted by 2BB models
have been selected in Paper I, based 
on the recent catalog \cite{Catalog20} 
presented also on the Web: \text{//www.
ioffe.ru/astro/NSG/thermal/cooldat.html}.
These can be cooling neutron stars with 
dipole magnetic fields and additionally heated
polar caps. This selection is illustrative
(does not pretend to be complete).

1. XMMU J172054.5$-$372652 is a neutron star that is 
probably associated with the SNR G350.1$-$0.3 
\cite{Gaensler2008-cco}; there is no direct evidence
of pulsations. 
\citet{Catalog20} used archival \chan\ data and fitted the X-ray
spectrum with a neutron star 
(\textsc{nsx}) atmosphere model assuming
$M = 1.4 \msun$ and $R = 13$ km. They obtained
$\Teff \approx 2$ MK, but did not perform
two-component fits which would be interesting.

2. PSR B1055$-$52 (J1057$-$5226) is a moderately magnetized middle-aged pulsar.
Its characteristic effective magnetic field, reported in the ATNF pulsar catalog \cite{ATNF}, is $\Beff=1.1 \times 10^{12}$~G.  
\citet{Catalog20} present the value 
$\kB T_\text{eff}^\infty \approx 70$ eV ($T_\text{eff}^\infty \approx 0.8$ MK) based on
the 2BB spectral fit by \citet{DeLuca_ea05}; the
fit includes also a power-law (PL)
non-thermal radiation component. \citet{Catalog20} have corrected the results of
\cite{DeLuca_ea05} for the distance estimate
to the source made by \citet{MignaniPK10} and reported $T_\text{effh}/T_\text{effc} \sim 2.3$
with $\shh \ll \scc$. It seems worth to try
to explain these data with a 2BB model containing hotspots.

3. PSR J1740+1000 possesses the magnetic field
$\Beff =1.8 \times 10^{12}$~G. 
The  2BB spectral fit was performed in \cite{Kargaltsev_ea12}. 
With the same version of the fit 
as selected in \cite{Catalog20}, one has
$\kB T_\text{eff}^\infty \approx 70$ eV ($T_\text{eff}^\infty \approx 0.8$ MK),
$T_\text{effh}/T_\text{effc} \sim 2.8$ and $\shh \ll \scc$,  
with the same
conclusion as for the PSR B1055$-$52.

4. PSR B1823$-$13 (J1826$-$1334), 
located in the SNR G18.0$-$00.7, 
has $\Beff=2.8 \times 10^{12}$~G.
Its X-ray emission is mainly non-thermal  \cite{PavlovKB08,Zhu_ea11} but has 
some thermal component. The 1BB+PL fit gives the radius
of thermally emitting region $\Reff \approx 5$ km, smaller than the expected radius
of a neutron star. Adding the second BB component is
statistically insignificant
with the present data, but might be possible in the future.

5. RX J1605.3+3249 (RBS 1556) is a  
neutron star studied by many authors (e.g. \cite{Motch_05,Posselt_ea07,Tetzlaff_ea12,Pires_ea19,
Malacaria_19}) with contradictory conclusions on its  properties (as detailed in \cite{Catalog20}). 
There is no solid evidence of stellar rotation
\cite{Pires_ea19}. Its
X-ray emission has been analyzed using
BB fits and neutron star atmosphere models.
Recently \citet{Pires_ea19}  analysed the \xmm\
observations and \citet{Malacaria_19}
jointly analysed the \textit{NICER} and
\textit{XMM-Newton} data. These teams improved
2BB fits by adding a broad Gaussian absorption line,
in which case they got $\kB T_{\rm effc}^\infty \sim 60$ eV ($T_{\rm effc}^\infty \sim 0.7$ MK),
and $T_{\rm effh}/T_{\rm effc} \sim 2$, a good opportunity 
to assume a dipole magnetic field and
additionally heated poles.

6. RX J1856.5$-$3754 is a neutron star with nearly thermal
spectrum. It was discovered by \citet{WalterWN96}. It is 
rotating with the spin period of $\sim$7 s; the effective magnetic field is 
$\Beff \sim 1.5 \times 10^{13}$~G; magnetic properties are 
highly debated (e.g. \cite{popov17,grandis21}
and references therein).
The spectrum has been measured  in a wide energy range,
including X-rays, optical and radio, and interpreted with
many spectral models, particularly, with the model of 
thin partially ionized magnetized hydrogen atmosphere 
on top of solidified iron surface (see, e.g., \cite{Ho_etal07,Potekhin14}).
Note alternative 2BB,
2BB+PL, and 3BB fits constructed by
\cite{Sartore_ea12} and \cite{Yoneyama_ea17}. 
The 2BB fits give $\kB T_{\rm effc}^\infty \sim 40$~eV 
($T_{\rm effc}^\infty \approx 0.46$~MK), 
$T_{\rm effh} /T_{\rm effc} \sim 1.6$, $R_{\rm effh} \sim 0.5 R$ and
$R_{\rm effc}\sim R$; they seem closer to the
2BB spectral models with smooth magnetic atmosphere, than 
2BB fits for other sources. Extra heating of magnetic
poles would simplify this interpretation.

This brief analysis does not reveal any
good candidate which would belong to 
the family of neutron stars 
with dipole magnetic fields and smooth surface
temperature distribution (Sect. \ref{s:acc}). This
does not mean that such neutron stars do not exist;
it can be difficult to identify them
because their spectra are close to 1BB spectrum.
To interpret the above sources one needs to
complicate the model, for instance, by introducing
extra heating of magnetic poles.

Now let us summarize our results. Following Paper I we have studied simple models 
(Sect. \ref{s:model}) 
of thermal spectra emitted from surfaces of
isolated neutron stars with dipole surface magnetic fields 
$10^{11} \lesssim \Bpole \lesssim 10^{14}$~G. 
Such fields make the surface temperature
distribution noticeably non-uniform. The model assumes BB emission  
with a local temperature from any surface element.

In Sect.\ \ref{s:acc} we have shown that the spectral 
X-ray fluxes emitted from such neutron stars are almost
the same for heat blaketing envelopes composed of iron and 
fully acctereted matter (at the same average
effective surface temperatures $\Teff$). By measuring
the spectral fluxes and $\Teff$, it would be difficult
to infer composition of their heat blanketing envelopes,
although this composition can noticeably affect neutron
star cooling (e.g., \cite{BPY21} and references therein).
In the presence of fully accreted matter, the fluxes are
accurately fitted by 2BB models (as in Paper I for the
iron heat blankets). At $M \sim 1.4\, \msun$ and
$R\sim 12$ km, the ratio $T_{\rm effh}/T_{\rm effc}$
of effective temperatures of the hotter to colder 
BB components cannot be essentially larger than $\sim 1.2$.
For less compact stars, with smaller $M/R$, this
value can be somewhat higher (see e.g. \cite{page95}, Paper I).
If this ratio is larger from observations, then the 
model of Sects. \ref{s:model} and
\ref{s:acc} cannot explain the data.

In Sect.\ \ref{s:hotspots}, following Paper I, we
have extended the model by assuming sufficiently weak
extra heating of one or both magnetic poles. Increasing
extra relative temperature rise $\delta$ at the pole(s), 
does not greatly violate the accuracy of the 2BB spectral  
approximation, but allows one to obtain larger
$T_{\rm effh}/T_{\rm effc}$. Also, it affects the pulse
fraction, which increases with $\Bpole$ and photon
energy $E$. The pulse fraction
can be very high if one magnetic pole is additionally
heated, while the other is not.  

Let us stress, that there have been many other
studies of thermal emission from magnetized
neutron stars. The emission produced by anisotropic
surface temperature distribution of the star with
dipole magnetic fields was analyzed by  \citet{page95}. 
Quadrupole magnetic field components
were added in \cite{page96}.
A possible presence of toroidal field component
in the neutron star crust
was studied in \cite{geppert2006,zane-turolla2006}. 
Also, there were studies of thermal emission
during magnetic
field evolution in
neutron stars (e.g., \cite{,popov17,grandis21}). These works give a
variety of magnetic fields configurations,
surface temperature distributions,
and phase-resolved spectra of neutron stars
to be compared with observations and
determine simultaneously both, the surface temperature
distribution and magnetic field geometry.

Introducing some extra heating of magnetic
poles, we are actually doing the same. Note that hot
spots can be produced by pulsar mechanism. Also, the
toriodal crustal magnetic fields can significantly widen the cold
equatorial belt, creating relatively stronger heated
poles \cite{geppert2006}, which can mimic hotspots. 
At the first step 
it might be simpler to
introduce phenomenological polar heating, that
can be different on the two poles, and to determine
the parameters of the extra heating 
(the $\Ts$ distribution) by varying
the parameters like $\delta$ and extra temperature
profile [Eq.\ (\ref{e:Ts})] at both poles. This
would be a simple and physically
transparent model of surface temperature distribution.
At the next step, one could infer (constrain) the
magnetic field geometry. This method can be easily extended
to different $\Ts$ distributions, not necessarily
axially symmetric.

Another very important direction would be to go beyond the
approximation of local BB emission from any surface patch,
and use more realistic emission models (with spectral
features and anisotropic radiation, as well
as with account of polarization effects,
e.g. \cite{HoPC08,ZKSP21}). 


\vspace{6pt} 




\funding{This research was partly (in analyzing the case of accreted matter) supported by the
Russian Science Foundation, grant 19-12-00133.}


	

\dataavailability{The data underlying this article will be shared on
	reasonable request to the corresponding author.} 

\acknowledgments{I am grateful to Oleg Gnedin for providing me photos of his Father.}




\abbreviations{Abbreviations}{
	The following abbreviations are used in this manuscript:\\
	BB -- blackbody}
\end{paracol}
\reftitle{References}

\externalbibliography{yes}

\newcommand{\aap}{Astron. Astrophys.}
\newcommand{\aj}{Astron. J.}
\newcommand{\apjl}{Astrophys. J. Lett.}
\newcommand{\apj}{Astrophys. J.}
\newcommand{\apjs}{Astrophys. J. Suppl. Ser.}
\newcommand{\apss}{Astrophys. Space Sci.}
\newcommand{\mnras}{Mon. Not. R. Astron. Soc.}
\newcommand{\nat}{Nature}
\newcommand{\pasa}{Publ. Astron. Soc. Australia}
\newcommand{\pasj}{Publ. Astron. Soc. Japan}
\newcommand{\pasp}{Publ. Astron. Soc. Pacific}
\newcommand{\prc}{Phys. Rev. C}
\newcommand{\pre}{Phys. Rev. E}
\newcommand{\prl}{Phys. Rev. Lett.}
\newcommand{\qjras}{Quarterly J. R. Astron. Soc.}
\newcommand{\sovast}{Sov. Astron.}
\newcommand{\ssr}{Space Sci. Rev.}



\label{lastpage}

\end{document}